\documentclass[9pt,twocolumn,twoside]{pnas-new}
\usepackage{esvect}
\usepackage{hyperref}
\usepackage{xspace}
\usepackage{mathtools}
\usepackage{wasysym}
\usepackage{stmaryrd}
\usepackage{marvosym}
\usepackage{stackengine}
\usepackage{setspace}
\usepackage{comment}
\newcommand{\steadyconstant}{steady state is the equal-probability
mixture\xspace}

\newcommand{\tmix}{t_{\text{mix}}}
\newcommand{\trel}{t_{\text{rel}}}

\newcommand{\ZERO}{\fbox{\phantom{$\bullet$}}}
\newcommand{\ONE}{\fbox{$\bullet$}}

\newcommand{\TWO}{\fbox{$\mathrlap{\mkern-4mu \rightarrow}\bullet$}}
\newcommand{\THREE}
{\fbox{$\mathrlap{\mkern-4mu\rightarrow} \mathrlap{\mkern-5mu 
\leftarrow}\bullet$}}
\newcommand{\FOUR}{\fbox{$\mathrlap{\mkern-5mu 
\leftarrow}\bullet$}}

\newcommand{\pit}[1]{\pi^{\{#1\}}}

\newcommand{\psiTwoOneOne}[3]{\psi_{\SET{\! \vv{#1} \!, #2, #3} \!}}
\newcommand{\psiOneTwoOne}[3]{\psi_{\SET{\! #1, \vv{#2}, #3} \!}}
\newcommand{\psiOneOneTwo}[3]{\psi_{\SET{\! #1, #2, \vv{#3}} \!}}
\newcommand{\eig}{E}

\newcommand{\ATwoOneOne}{A_{\! \vv{\bullet} \! \circ \circ}}
\newcommand{\AOneTwoOne}{A_{\circ \!  \vv{\bullet}  \! \circ}}
\newcommand{\AOneOneTwo}{A_{\circ \circ \! \vv{\bullet} \! }}
\newcommand{\BTwoOneOne}{B_{\! \vv{\bullet} \! \circ \circ}}
\newcommand{\BOneTwoOne}{B_{\circ \!  \vv{\bullet}  \! \circ}}
\newcommand{\BOneOneTwo}{B_{\circ \circ \! \vv{\bullet} \! }}
\newcommand{\CTwoOneOne}{C_{\! \vv{\bullet} \! \circ \circ}}
\newcommand{\COneTwoOne}{C_{\circ \!  \vv{\bullet}  \! \circ}}

\newcommand{\DTwoOneOne}{D_{\! \vv{\bullet} \! \circ \circ}}
\newcommand{\DOneTwoOne}{D_{\circ \!  \vv{\bullet}  \! \circ}}

\newcommand{\ETwoOneOne}{E_{\! \vv{\bullet} \! \circ \circ}}
\newcommand{\EOneTwoOne}{E_{\circ \!  \vv{\bullet}  \! \circ}}

\newcommand{\FTwoOneOne}{F_{\! \vv{\bullet} \! \circ \circ}}
\newcommand{\FOneTwoOne}{F_{\circ \!  \vv{\bullet}  \! \circ}}
\newcommand{\FOneOneTwo}{F_{\circ \circ \! \vv{\bullet} \! }}

\newcommand{\OmegaRW}{\Omega^{\text{\textsc{RW}}}}
\newcommand{\LRW}{\LCAL^{\text{\textsc{RW}}}}

\newcommand{\OmegalRW}{\Omega^{\text{l-\textsc{RW}}}}
\newcommand{\OmegaSSEP}{\Omega^{\SSEP}}
\newcommand{\OmegaTASEP}{\Omega^{\TASEP}}
\newcommand{\OmegalTASEP}{\Omega^{\text{l-\TASEP}}}

\newcommand{\alphacrit}{\alpha_\text{crit}}
\newcommand{\tauauto}{\tau_{S(q)}}
\newcommand{\LCALtilde}{\tilde{\LCAL}}

\newcommand{\dTV}{d_\text{TV}}
\newcommand{\RW}{\textsc{RW}}
\newcommand{\lRW}{\text{L-RW}}
\newcommand{\TASEP}{\textsc{Tasep}\xspace}
\newcommand{\lTASEP}{\textsc{l-T}\xspace}
\newcommand{\SSEP}{\textsc{Ssep}\xspace}
\newcommand{\pullback}{pullback\xspace}
\newcommand{\oo}{\cdot}
\newcommand{\be}{\begin{equation}}
\newcommand{\ee}{\end{equation}}
\newcommand{\bea}{\begin{eqnarray}}
\newcommand{\eea}{\end{eqnarray}}

\newcommand{\Perm}{Q}

\def\nn{\nonumber\\}
\def\fr#1{[\ref{#1}]}
\def\EV{E}

\templatetype{pnasresearcharticle} 

\newcommand\subfig[2]{{Fig.~\ref{#1}{#2}}}
\newcommand\subcap[1]{{(#1):}}

\newcommand{\SET}[1]{\{#1\}}

\newcommand{\eq}[1]{Eq.~[\ref{#1}]}
\newcommand{\eqtwo}[2]{Eqs~[\ref{#1}] and~[\ref{#2}]}

\newcommand{\fig}[1]{Fig.~\ref{#1}}

\newcommand{\quot}[1]{``#1''}

\newcommand{\eg}{\textrm{e.g.}}
\newcommand{\cf}{\textrm{cf}}

\newcommand{\ie}{\textrm{i.e.}}

\newcommand{\LCAL}{\mathcal{L}}  
\newcommand{\alphabar}{\overline{\alpha}}  
\newcommand{\expa}[1]{\mathrm{e}^{#1}}   
\newcommand{\expb}[1]{\exp \glb #1 \grb} 
\newcommand{\loga}[2][]{\log^{#1}\! \gla #2 \gra}  

\newcommand\Var{\operatorname{Var}}

\newcommand{\gla}{\,}  
\newcommand{\gra}{}  
\newcommand{\glb}{\left(}  
\newcommand{\grb}{\right)}  
\newcommand{\gld}{\left\{}  
\newcommand{\grd}{\right\}}  
\newcommand{\gle}{\left|}  
\newcommand{\gre}{\right|}  

\newcommand{\const}{\text{const}}

\newcommand{\TO}{,\ldots,}

\newcommand{\Phat}{\widehat{P}}

\newcommand{\pihat}{\hat{\pi}}
\newcommand{\Pihat}{\hat{\Pi}}

\newcommand{\Omegahat}{\widehat{\Omega}}

\newcommand{\half}{\frac{1}{2}}

\title{Lifted TASEP: a Bethe ansatz integrable paradigm for  
non-reversible Markov chains} 

\author[a,b]{Fabian H. L.  Essler}
\author[b]{Werner Krauth}

\affil[a]{Rudolf Peierls Centre for Theoretical Physics, Clarendon Laboratory, 
Oxford OX1 3PU, UK}
\affil[b]{Laboratoire de Physique de l’Ecole normale sup\'erieure, ENS, 
Universit\'e PSL, CNRS, Sorbonne Universit\'e, Universit\'e de Paris Cit\'e, 
24 rue Lhomond, 75005 Paris, France}

\leadauthor{Essler}

\begin{abstract}
Markov-chain Monte Carlo (MCMC), the field of stochastic algorithms built on the
concept of sampling, has countless applications in science and technology. 
The overwhelming majority of MCMC algorithms are time-reversible and
satisfy the detailed-balance condition, just like physical systems in
thermal equilibrium. The underlying Markov chains typically display
diffusive dynamics, which leads to a slow exploration of sample
space. Significant speed-ups can be achieved by non-reversible MCMC
algorithms exhibiting non-equilibrium dynamics, whose
steady states exactly reproduce the target equilibrium
states of reversible Markov chains. Such algorithms have had
successes in applications but are generally difficult to analyze,
resulting in a scarcity of exact results. Here, we introduce the
\quot{lifted} \TASEP (totally asymmetric simple exclusion process) as
a paradigm for lifted non-reversible Markov chains. Our model can be viewed as a
second-generation lifting of the reversible Metropolis algorithm on a
one-dimensional lattice and is exactly solvable by an unusual kind of
coordinate Bethe ansatz. We establish the integrability of the model
and present strong evidence that the lifting leads to faster
relaxation than in the KPZ (Kardar--Parisi--Zhang) universality class.
\end{abstract}

\dates{This manuscript was compiled on \today}

\begin{document}
\def\dosquare{\kern-.1em\tikz[overlay]\draw[red,rounded
corners,fill=red!15!white,baseline] (0,1.9ex) rectangle
(2.4em,-3.2ex);\kern.1em}

\maketitle
\fancyfoot[LE]{\footerfont\textbf{\thepage}}
\fancyfoot[RO]{\footerfont\textbf{\thepage}}
\fancyfoot[RE,LO]{}
\ifthenelse{\boolean{shortarticle}}{\ifthenelse{\boolean{singlecolumn}}{
\abscontentformatted}{\abscontent}}{}

\dropcap{E}ver since its beginnings~\cite{Metropolis1953} in 1953, the field of
Markov-chain Monte Carlo has relied on the concept of reversibility to
sample probability distributions $\pi$ in high-dimensional sample spaces
$\Omega$. Starting from an initial configuration $x_0$, elements in the chain
$x_0, x_1, \dots, x_t,x_{t+1} \dots$ sample $\pi$ for large Monte-Carlo times
$t$. In a reversible Markov chain at equilibrium (at large $t$), the chain
link $\dots, x_t, x_{t+1}, \dots$ appears with the same probability as
the time-reversed link $\dots, x_{t+1}, x_{t}, \dots$.
The probability $P(x_t,x_{t+1})$ of a transition from $x_t$ to
$x_{t+1}$ thus fulfils the famous detailed-balance condition
  $\pi_{x_t} P(x_t, x_{t+1}) = \pi_{x_{t+1}} P(x_{t+1}, x_t)$.
Reversible Markov chains like the Metropolis~\cite{Metropolis1953} and
the heatbath~\cite{Glauber1963,Creutz1980HeatBath,Geman1984}
algorithms mimic physical systems in thermal equilibrium, where all
net flows vanish, so that the dynamics is diffusive and hence slow.

Markov chains need not be reversible, as a steady state $\pi$
characterized by the \quot{global-balance} condition $\pi_x = \sum_{y \in
  \Omega} \pi_y P_{y x}$ can be targeted without respecting detailed
balance. The resulting non-reversible Markov chains can be viewed as
out-of-equilibrium processes that approach the equilibrium
distribution $\pi$ for large $t$. Their non-vanishing flows often
allow them to relax faster than diffusively. However, for many
decades, genuinely non-reversible Markov chains that relax to a target
equilibrium distribution were hard to construct. This changed with the
advent of lifted Markov chains~\cite{Diaconis2000,Chen1999}. 

Lifted Markov chains can be constructed from reversible chains as illustrated
by the random walk on the lattice $\OmegaRW = \SET{1 \TO L}$ and its set of
moves $\LRW = \SET{-1,+1}$ (see Methods for definitions). At each
time $t=0,1,\dots$, a random move $\sigma$ (forward or backward) is
applied to $x_t$ and results in a new sample $x_{t+1}$: 
\begin{equation}
\underbrace{\overset{1}{\ZERO} \ZERO \ONE \overset{L}{\ZERO}}_{ x_t \in 
\OmegaRW} 
 \rightarrow 
\underbrace{\ZERO \ZERO \THREE \ZERO}_{x_t
\in \OmegaRW, \sigma \in
\LRW}  \rightarrow 
\underbrace{
\begin{cases}
\ZERO \ONE \ZERO \ZERO &p = \half\\
\ZERO \ZERO \ZERO \ONE & p = \half
\end{cases}}_{x_{t+1}\in \OmegaRW}.
\label{equ:RandomWalkDef1}
\end{equation}
Moves across the end points are rejected:
\begin{equation}
\underbrace{\overset{1}{\ONE} \ZERO \ZERO \overset{L}{\ZERO}}_{x_t =1 \text{ or 
} x_t= L}  
\rightarrow
\underbrace{\THREE \ZERO \ZERO \ZERO}_{\text{\!\!\!rejected back move\!\!\!}}  
\rightarrow 
\underbrace{
\begin{cases}
\ONE \ZERO \ZERO \ZERO& p= \half\\
\ZERO \ONE \ZERO \ZERO& p= \half\\
\end{cases}}_{x_{t+1}\in \OmegaRW}.
\label{equ:RandomWalkDef2}
\end{equation}
The random walk is reversible with equal stationary probabilities
$\pi^\RW(x) = \tfrac1L$ for all $x$, but takes $\sim L^2$ moves to
cover all $L$ sites, signaling diffusive behavior.

The \emph{lifted} random walk \quot{duplicates} each $x_t \in
\OmegaRW$
into an enlarged sample space $\OmegalRW = \OmegaRW \times \LRW$
\begin{equation}
\underbrace{\ZERO \ZERO \overset{i}{\ONE} \ZERO}_{x_t \in \OmegaRW}
\rightarrow
\underbrace{
\begin{cases}
\ZERO \ZERO \overset{\mathrlap{\mkern-18mu (i,\sigma)}}{\TWO} \ZERO \\
\ZERO \ZERO \FOUR \ZERO
\end{cases}}_{x_t \in \OmegalRW},
\label{equ:LiftingRW}
\end{equation}
which now contains $2 L$ elements.
In the bulk of the system, the lifted random walk first moves from $(i,\sigma)$
to $(i+ \sigma, \sigma)$, then,  with probability
$\alpha=1-\bar{\alpha}$, flips the direction from $\sigma$ to $-\sigma$:
\begin{align}
\underbrace{\ZERO \ZERO \overset{i}{\FOUR} \ZERO}_{x_t \in \OmegalRW} 
&\xrightarrow{p=1} 
\ZERO \overset{\!\!i+\sigma\!\!}{\FOUR} \ZERO \ZERO  
\rightarrow 
\underbrace{
\begin{cases}
\ZERO \TWO \ZERO \ZERO  & p = \alpha\\
\ZERO \FOUR \ZERO \ZERO  & p = \alphabar
\end{cases}}_{x_{t+1} \in \OmegalRW, \alpha \ll 1,\alphabar = 1-\alpha}, 
\label{equ:LiftedRandomWalk1}
\intertext{
while at a boundary it may only flip $\sigma$:}
\underbrace{\FOUR \ZERO \ZERO \ZERO}_{\text{$x_t$ at end point}
} &\xrightarrow{p=1} 
\underbrace{\TWO \ZERO \ZERO \ZERO}_{\sigma \to -\sigma} \rightarrow 
\underbrace{
\begin{cases}
\FOUR \ZERO \ZERO \ZERO  & p = \alpha\\
\TWO \ZERO \ZERO \ZERO  & p =  \alphabar
\end{cases}}_{x_{t+1} \in \Omega^\lRW}.
\end{align}
The lifted random walk is non-reversible (the move from $(i, \sigma)$ to
$(i+\sigma, \sigma)$ cannot be reversed), but its
\steadyconstant
of all $2L$ lifted
configurations. For $\alpha = 1/L$, it moves in $\sim L$ steps
through the lifted lattice of $2L$ sites, much faster than the 
original random walk. The increased sampling speed in the
augmented space $\OmegalRW$ also boosts the exploration of the original
$\OmegaRW$.

In the present work, we consider liftings of the Metropolis algorithm for hard
spheres a one-dimensional lattice, which is nothing but the symmetric simple
exclusion process (\SSEP)
\cite{Spitzer1970,Lacoin2016detailed,Lacoin_2017_SSEP}. We introduce a
second-generation non-reversible lifting, the lifted \TASEP (totally asymmetric
simple exclusion process), that again speeds up the timescales on which sample
space is explored. The lifted \TASEP has two striking properties. First, it
features non-local moves, yet it can be solved exactly by the Bethe ansatz. This
allows us us to compute exact eigenstates of the transition matrix for large
system sizes. Our analysis suggests that it belongs to a new class of Bethe
ansatz solvable models with relaxation rates that are faster than for the KPZ
(Kardar--Parisi--Zhang) class. Second, as our construction is based on the
lifting principle~\cite{Diaconis2000,Chen1999}, our model immediately
generalizes to the continuum~\cite{KapferKrauth2017,Lei2019}. It also connects
to  lifted MCMC algorithms in higher dimensions~\cite{Bernard2009,Maggs2022},
and with smooth interactions~\cite{Peters_2012,Michel2014JCP}, which have been
successfully applied in statistical mechanics~\cite{Bernard2011,Kampmann2021}
and chemical physics~\cite{Faulkner2018}.

\section*{Exclusion models: from the \SSEP to the lifted \TASEP}

The \SSEP implements the reversible Metropolis algorithm for indistinguishable
hard spheres on an $L$-site lattice with periodic boundary conditions. A known
non-reversible lifting of the SSEP (with periodic boundary conditions) is the
much studied
\TASEP~\cite{Gwa1992six,Gwa1992bethe,Kim1995Bethe,derrida1998exactly,
schutz2001exactly,ChouTASEP2011}. Here we introduce a \quot{second-generation}
lifting of the SSEP by constructing a lifting of the \TASEP. It
implements the same strategy for shortening relaxation times as the
lifted random walk discussed in the introduction.

\subsection*{Symmetric simple exclusion process}
In the \SSEP, configurations are described by the occupied sites $\SET{\dots i,
j, k \dots}$. At each time step, a randomly chosen active particle
\begin{equation}
\underbrace{\ZERO \overset{i}{\ONE} \overset{j}{\ONE} \ZERO
\overset{k}{\ONE}}_{x_t \in
\Omega^\SSEP} \rightarrow
\underbrace{
\begin{cases}
\ZERO \overset{i}{\THREE} \overset{j}{\ONE} \ZERO \overset{k}{\ONE}& p = 
\tfrac1N\\
\ZERO \ONE \THREE \ZERO \ONE& p = \tfrac1N\\
\ZERO \ONE \ONE \ZERO \THREE& p = \tfrac1N
\end{cases}}_{\text{choice of active particle}
} \rightarrow \cdots,
\label{equ:SSEPDef1}
\end{equation}
attempts a random forward or backward move:
\begin{equation}
\cdots \rightarrow
\underbrace{
\ZERO \THREE \ONE \ZERO \ONE}_{\text{choice of move}
} \rightarrow
\underbrace{ \begin{cases}
\ONE \ZERO \ONE \ZERO \ONE & p= \tfrac12 \\
\ZERO \ONE \ONE \ZERO \ONE & p= \tfrac12
\end{cases} }_{x_{t+1} \in \Omega^\SSEP}\ .
\label{equ:SSEPDef2}
\end{equation}
Moves that would violate the exclusion condition (at most one particle
per site) are rejected, resulting in $x_{t+1}=x_t$. The \SSEP is
reversible and its \steadyconstant of all
configurations $\pi_x = \const\quad \forall x \in \Omega^\SSEP$
(see Methods for examples). The relaxation time (see Methods)
scales as $\trel^\SSEP \sim N^3$. In contrast,
the approach towards equilibrium from a worst-case initial
configuration takes place on a
non-asymptotic~\cite{AldousDiaconis1986} mixing time scale $\tmix$,
which scales as $\sim N^3 \log N$, i.e. a factor $\log N$ larger
than the relaxation time. The approach to equilibrium thus takes longer than
the de-correlation in equilibrium, giving rise to the celebrated \quot{cutoff
phenomenon}~\cite{AldousDiaconis1986,Lacoin2016detailed,Lacoin_2017_SSEP},
a sudden convergence towards $\pi$ at $t \sim \tmix$.

\subsection*{Totally symmetric simple exclusion process}
One possible lifting of the \SSEP
is the \quot{bi-directional} \TASEP, where the sample space is augmented
$\OmegaTASEP =
\OmegaSSEP \times \SET{-1,+1}$ while the move set is reduced
(see Methods for an example transition matrix of the
bi-directional \TASEP). For
periodic boundary conditions  we may restrict ourselves to the forward-moving
sector. Then, at each time step $t=0, 1, \dots$, a randomly chosen active
particle
\begin{equation}
\overrightarrow{
\underbrace{\underset{1}{\ZERO} \ONE \ONE \ZERO \underset{L}{\ONE}}_{x_t \in 
\Omega^\TASEP}} \rightarrow
\begin{cases}
\ZERO \TWO \ONE \ZERO \ONE \rightarrow 
\overrightarrow{\ZERO \ONE \ONE \ZERO 
\ONE}  \\
\ZERO \ONE \TWO \ZERO \ONE \rightarrow \overrightarrow{\ZERO \ONE \ZERO \ONE
\ONE}  \\
\underbrace{\ZERO \ONE \ONE \ZERO \underset{L}{\TWO}}_{\text{active particle}} 
\rightarrow \overrightarrow{\underbrace{\underset{\mathrlap{\mkern-25mu L+1 
\equiv 1}}{\ONE} \ONE \ONE \ZERO 
\ZERO}_{x_{t+1} \in \Omega^\TASEP} } \\
\end{cases}
\label{equ:MoveTASEP}
\end{equation}
attempts a forward move, which is rejected if it violates the exclusion
condition. The \TASEP is non-reversible, and its
\steadyconstant
of all configurations (see Methods for
an example). In the \TASEP, both the relaxation time~\cite{Dhar1987,Gwa1992bethe}
and the mixing time~\cite{BaikLiu2016} scale as $\trel, \tmix \sim N^{5/2}$,
i.e.
much faster than for the \SSEP. The exponent 
$5/2$ reduces to $3/2$ if there are $\sim N$ moves per unit time (see
Methods) which is characteristic of the KPZ universality
class \cite{Kardar1985dynamic}.

\subsection*{Lifted \TASEP}
\label{ssec:LTASEP}
In the lifted \TASEP, a configuration is characterized by the positions of the
particles, and in addition the identity of the active particle (and in
principle its direction, see Methods). Restricting ourselves to the
forward-moving sector, a lifted-\TASEP configuration is denoted by
$\SET{\dots j,\vv{k},l \dots}$ and refers to a three-state model,
where sites are either empty, occupied with a \quot{regular} article,
or with the active particle, which is the only one that can move. 
Specifically, the deterministic first part of the move
\begin{align}
\underbrace{\ZERO \overset{j}{\ONE} \overset{\vv{k}}{\TWO} \ZERO
\overset{l}{\ONE}}_{x_t \in \Omega^{\text{l-\TASEP}}}
&\rightarrow
\underbrace{
\ZERO \overset{j}{\ONE} \ZERO \overset{\!\!\!\vv{k+1}\!\!\!}{\TWO}
\overset{l}{\ONE}}_{\mathrlap{\mkern-90mu \text{\quot{particle}
displacement}} }
\xrightarrow{\text{\eq{equ:firstMickey}b}} \cdots
\label{equ:firstMickey} \\
\underbrace{
\ZERO \overset{j}{\ONE} \overset{\vv{k}}{\TWO} \overset{\!\!\!k+1\!\!\!}{\ONE}
\ZERO}_{x_t \in \Omega^{\text{l-\TASEP}}} &\rightarrow
\underbrace{
\ZERO \overset{j}{\ONE} \overset{k}{\ONE} \overset{\!\!\!\vv{k+1}\!\!\!}{\TWO}
\ZERO}_{\text{\quot{lifting} move
}}
\xrightarrow{\text{\eq{equ:secondMickey}b}} \cdots
\label{equ:secondMickey}
\end{align}
is complemented by a stochastic \quot{\pullback} with probability $\alpha$,
which
transfers the
\quot{activity} from the active particle to the particle
preceding it, without modifying the set of positions:
\begin{align}
\cdots \rightarrow
\ZERO \overset{j}{\ONE} \ZERO \overset{\!\!\!\vv{k+1}\!\!\!}{\TWO}
\overset{l}{\ONE}   & \rightarrow
\begin{cases}
 \ZERO \overset{\vv{\jmath}}{\TWO} \ZERO \overset{\!\!\!k+1\!\!\!}{\ONE}
\overset{l}{\ONE} \quad &p=\alpha \\
\ZERO \underset{j}{\ONE} \ZERO \underset{\!\!\!\vv{k+1}\!\!\!}{\TWO}
\underset{l}{\ONE} \quad &p=\alphabar
\end{cases}
\tag{\ref{equ:firstMickey}b} \\
\cdots \rightarrow
\ZERO \ONE \ONE \TWO  \ZERO  &\rightarrow 
\underbrace{
\begin{cases}
\ZERO {\ONE} {\TWO} \ONE \ZERO \quad &p=\alpha
\\
\ZERO \ONE \ONE \TWO \ZERO \quad &p= \alphabar
\end{cases}}_{x_{t+1} \in \Omega^{\lTASEP} }
\tag{\ref{equ:secondMickey}b}
\end{align}
The \pullback can be very non-local, as for  example in the
move $\SET{\dots j,\vv{k+1},l \dots} \to
\SET{\dots \vv{\jmath},k+1,l \dots}$ of \eq{equ:firstMickey}b
for $j\ll k$.
  
The non-zero \emph{row elements} of the transition matrix
$P^\lTASEP$ can be read off from
\eqtwo{equ:firstMickey}{equ:secondMickey}.
For $k+1 < l$, we have
\begin{equation}
\begin{aligned}
P^\lTASEP(\SET{\dots j , \vv{k},l  \dots}, \SET{\dots j , \vv{k+1},l \dots })
&= \alphabar,
\\
P^\lTASEP(\SET{\dots j , \vv{k},l  \dots}, \SET{\dots \vv{\jmath}, k+1,l
\dots}) &=
\alpha,
\end{aligned}
\label{equ:RowTransitionMatrixFar}
\end{equation}
while for $k+1 = l$, the non-zero elements are
\begin{equation}
\begin{aligned}
P^\lTASEP(\SET{\dots j , \vv{k},k+1  \dots}, \SET{\dots j , k, \vv{k+1} \dots
})
&= \alphabar,
\\
P^\lTASEP(\SET{\dots j , \vv{k},k+1  \dots}, \SET{\dots j, \vv{k}, k+1 \dots})
&=
\alpha.
\label{equ:RowTransitionMatrixClose}
\end{aligned}
\end{equation}
In the above, periodic boundary conditions are understood.
The two non-zero elements in
\eqtwo{equ:RowTransitionMatrixFar}{equ:RowTransitionMatrixClose} define the
\quot{$\alpha$}-move and the \quot{$\alphabar$}-move out of the
configuration $\SET{\dots j, \vv{k}, l \dots}$. From these two equations, one
may directly write down eigenvalue equations for the \emph{right}
eigenvectors of the transition matrix, namely, for $k+1 > l$:
\begin{align}
 \eig \psi^r_{\SET{\dots j , \vv{k},l  \dots}} &=
 \alpha \psi^r_{\SET{\dots \vv{\jmath}, k+1,l \dots}} +
 \alphabar \psi^r _{\SET{\dots j , \vv{k+1},l \dots }}).\\
 \intertext{and for $k+1 = l$:}
 \eig \psi^r_{\SET{\dots j , \vv{k},k+1  \dots}} &=
 \alpha \psi^r_{\SET{\dots j, \vv{k}, k+1 \dots}} +
 \alphabar \psi^r _{\SET{\dots j , k,  \vv{k+1} \dots }}).
\end{align}
Clearly, these are solved for $\eig=1$ by $\psi^r = \const$, so that
$P^\lTASEP$ is a valid stochastic matrix with unit row sums.
The transition matrix has non-zero diagonal elements
(see \eq{equ:RowTransitionMatrixClose}), so that it is aperiodic for
any \pullback $0<\alpha <1$.
The non-zero \emph{column elements} of the transition matrix are as follows.

\noindent
For $j < k-1$ and $l > k+1$:
\begin{equation}
\begin{aligned}
P^\lTASEP(\SET{\dots j , \vv{k-1},l  \dots}, \SET{\dots j , \vv{k},l \dots })
&= \alphabar, \\
P^\lTASEP(\SET{\dots j , k,\vv{l-1}  \dots}, \SET{\dots j, \vv{k},l
\dots}) &= \alpha\ .
\end{aligned}
\label{equ:LiftedTasepTransition1}
\end{equation}
For $j = k-1$ and $l > k+1$:
\begin{equation}
\begin{aligned}
P^\lTASEP(\SET{\dots \vv{k-1}, k,l  \dots}, \SET{\dots j , \vv{k},l \dots })
&= \alphabar, \\
P^\lTASEP(\SET{\dots k-1 , k,\vv{l-1}  \dots}, \SET{\dots j, \vv{k},l
\dots}) &= \alpha\ .
\end{aligned}
\label{equ:LiftedTasepTransition2}
\end{equation}
For $j < k-1$ and $l = k+1$:
\begin{equation}
\begin{aligned}
P^\lTASEP(\SET{\dots j, \vv{k-1},l  \dots}, \SET{\dots j , \vv{k},l
\dots }) &= \alphabar, \\
P^\lTASEP(\SET{\dots j , \vv{k},l  \dots}, \SET{\dots j, \vv{k},l
\dots}) &= \alpha\ .
\end{aligned}
\label{equ:LiftedTasepTransition3}
\end{equation}
For $j = k-1$ and $l = k+1$:
\begin{equation}
\begin{aligned}
P^\lTASEP(\SET{\dots \vv{j}, k,l  \dots}, \SET{\dots j , \vv{k},l
\dots }) &= \alphabar, \\
P^\lTASEP(\SET{\dots j , \vv{k},l  \dots}, \SET{\dots j, \vv{k},l
\dots}) &= \alpha\ .
\label{equ:LiftedTasepTransition4}
\end{aligned}
\end{equation}
The column elements of the transition matrix thus add up to one, so that
$P^\lTASEP$ is in fact \emph{doubly stochastic}. This implies that the
\steadyconstant of all configurations for $0< \alpha < 1$ and shows that the
model is indeed a lifting of the \TASEP. Furthermore, from any of these cases,
one may directly write down eigenvalue equations $\eig \psi = \psi P^\lTASEP$
for the \emph{left} eigenvectors $\psi$ of $P^\lTASEP$,
which we will further analyze using the Bethe ansatz. Again, $\psi = \pi =
\const$ is the left eigenvector with eigenvalue $1$, as already follows from
the double stochasticity.

The lifted \TASEP differs from the integrable defect particle in
the asymmetric exclusion process studied in Ref.~\cite{Derrida_1999} in two
important ways. First,  in our stochastic process only the defect particle
(that is, the active particle) moves. Second, our dynamics is
non-local because the stochastic \pullback interaction acts between
the closest particles, but they may be far separated in space.

\section*{Lifted \TASEP: basic properties}
While the value of the  \pullback $0 < \alpha < 1$ does not affect the steady
state, it plays a remarkable role in the dynamics of the lifted \TASEP. Any
lifted configuration can evolve either through the \quot{$\alpha$}-move or the
\quot{$\alphabar$}-move (see
\eqtwo{equ:RowTransitionMatrixFar}{equ:RowTransitionMatrixClose}). An
$\alphabar$-move advances the activity by one site. A series of $L-1$ sequential
$\alphabar$-moves translates the entire configuration backwards by one site. An
$\alpha$-move, on average over all lifted configurations, pulls the activity
back by $(L-N)/N$ sites. Averaging over $\alpha$-moves and $\alphabar$-moves
shows that the activity drifts with velocity $-\alpha L/ N + 1$, so that the
drift velocity vanishes
for $\alphacrit = N / L$. In event-chain Monte Carlo, the drift velocity of the
activity
is proportional to the system pressure~\cite{Michel2014JCP} and in the
one-dimensional harmonic model, the zero-pressure point exhibits fast
autocorrelation functions~\cite{Lei2018_TOP}. This behavior was numerically
confirmed in a continuous-space variant of the lifted \TASEP~\cite{Lei2019} (see
also Ref.~\cite{Maggs2022}). Our \pullback similarly generates a zero-pressure
point at $\alphacrit$  without affecting the steady-state properties.

\begin{figure}
\centering
\includegraphics[
width=0.8\linewidth]{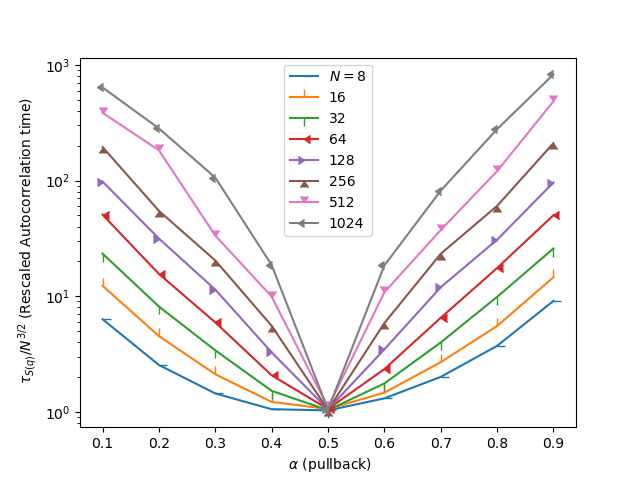}
\caption{Equilibrium autocorrelation time $\tauauto$ for the structure factor
$S(q)$ with $q = 2 \pi / L$ (see \eq{equ:StructureFactor}), as a function of the
\pullback $\alpha$, for different system sizes.}
\label{fig:AutocorrelationsAlpha}
\end{figure}
MCMC simulations for the lifted \TASEP confirm
the special role of the \pullback. We have analyzed
the autocorrelation time $\tauauto$ of the structure factor
in equilibrium~\cite{Sokal1997},
\begin{equation}
S(q,t) = \frac{1}{N} \gle \sum_{\xi \in \SET{\dots,i,j,k\dots}}  \expa{i q
\xi} \gre,
\label{equ:StructureFactor}
\end{equation}
where $q = 2 \pi / L $ is
the smallest wave number. The structure factor is sensitive to long-range
density fluctuations, which are expected to relax slowly in equilibrium. As a
function of $\alpha$, $\tauauto$ exhibits a pronounced dip at $\alpha =
\alphacrit$ (see \fig{fig:AutocorrelationsAlpha}).

\begin{figure}
\centering
\includegraphics[
width=1.0\linewidth]{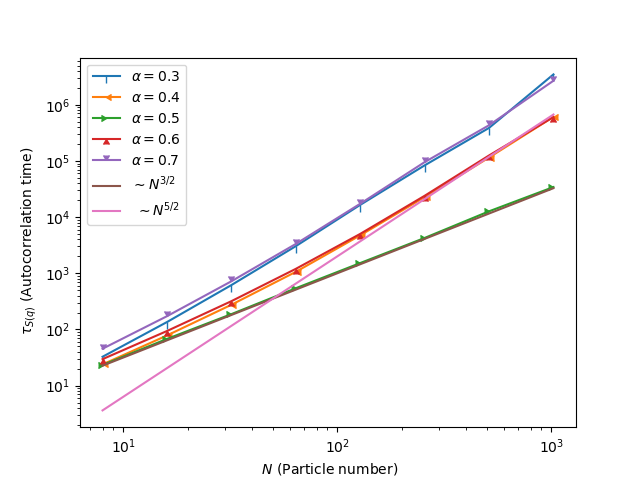}
\caption{Equilibrium autocorrelation time $\tauauto$ for the structure factor
$S(q)$ with $q = 2 \pi / L$, as a function of $N=L/2$, for
different values of the \pullback. }
\label{fig:AutocorrelationsN}
\end{figure}

At fixed $\alpha$, we observe a power-law behavior of $\tauauto$
as a function of  $N$, but with different exponents at $\alphacrit$ and away
from it (see also Ref.~\cite{Lei2019}). Taking $L = 2N$ we find that for $\alpha =
\alphacrit = \half$, these equilibrium autocorrelations
are compatible with  $\tauauto \sim N^{3/2}$, while for
non-critical $\alpha \in \SET{0.3, 0.4, 0.6, 0.7}$,
we observe $\tauauto \sim N^{5/2} $
(see \fig{fig:AutocorrelationsN}).
At $\alphacrit$, the motion of the activity  is
driftless but superdiffusive (see also Ref.~\cite{Lei2019}):
it only takes $\sim L^{3/2}$ steps for the activity to move through the
system of $L$ sites.

\begin{figure}[htbp]
\centering
\includegraphics[width=0.8 \linewidth]{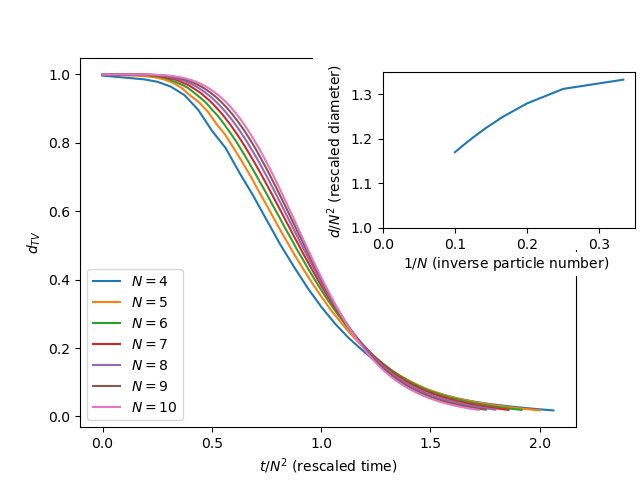}
\caption{Worst-case total variation distance $d_{\text{TV}}$
for  the lifted \TASEP
as a function of rescaled time $t$ for
small system sizes with $L= 2N$, at \pullback $\alpha = \alphacrit = \half$.
Inset: Diameter of the lifted \TASEP,
compatible with a scaling $\sim N^2$.
}
\label{fig:TVDLiftedTASEP}
\end{figure}

To estimate the mixing time $\tmix$ for $ \alpha= \alphacrit = \half$, we have
taken the compact state $ x_{t=0} = \SET{\vv{1},2 \TO N}$ as our conjectured
worst-case initial configuration, and computed in a numerically exact way the
total variation distance from the steady state $\pi$ (see Methods). Our data are
compatible with $\tmix \sim N^2$. This is a clear indication that the lifted
\TASEP mixes much faster than the \TASEP and does not belong to the KPZ
universality class. We note again that,  in our time units, the KPZ exponent is
$5/2$.
We have also estimated the diameter (the minimal time to
connect any two configuration $x , y \in \OmegalTASEP$) appears to scale as
$\sim N^2$ with a coefficient close to one (see \fig{fig:TVDLiftedTASEP}).

\subsection*{Spectrum of the transition matrix}
\begin{figure}[ht]
\centering
\includegraphics[width=\linewidth]{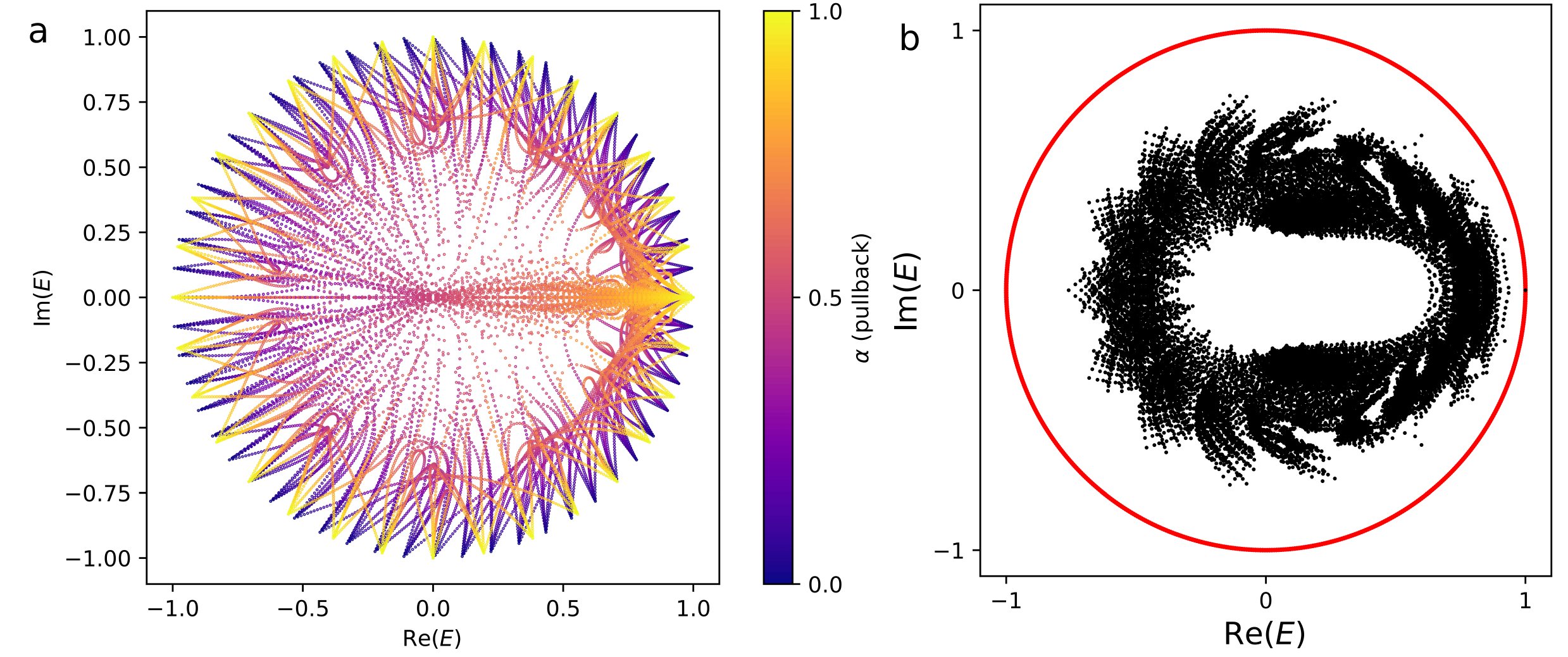}
\caption{Eigenvalue spectrum of the lifted \TASEP.
\subcap{a} Spectrum
for $N=4$, $L=8$
as a function of the \pullback $\alpha$. For $\alpha=0$ and $1$, all
eigenvalues $E$ are equally spaced on the complex unit circle.
\subcap{b} Spectrum for $N=7$, $L=14$ for $\alpha=\alphacrit=\half$, with a
single eigenvalue on the complex
unit circle. }
\label{fig:TASEPApplication}
\end{figure}

To illustrate the spectrum of the transition matrix $P^\lTASEP$ and its
dependence on the \pullback $\alpha$, we show in
\fig{fig:TASEPApplication} the complex eigenvalues for
$N=4, L=8$ as functions of $\alpha$.

For $L =2 N$
and  $\alpha=0$, the $N \binom{2N}{N}$ eigenvalues lie on $2N(2N-1)$
equally spaced points on the complex unit circle.
Each of these eigenvalues has degeneracy
\begin{equation*}
C(N-1) = N \binom{2N}{N} / [2N(2N-1)],
\end{equation*}
where $C$ is the Catalan number. For $\alpha =1$, the eigenvalues lie on
$2 N^2$
equally spaced points on the complex unit circle, and all the
eigenvalues except $\eig = (1,0)$ are again $C(N-1)$ times degenerate, while
the eigenvalue $\eig=(1,0)$ has a degeneracy of $\binom{2N}{N} - (2N^2 -1)
C(N-1)$ (see \subfig{fig:TASEPApplication}{a}).
Clearly, in these two cases, the transition matrix
$P^\lTASEP$ is not
irreducible (except for $N=2$, $\alpha = 0$), and it is periodic for all $N$.
In contrast,
for $0<\alpha <1$, the transition matrix $P^\lTASEP $ has only a single
eigenvalue with $|E|=1$ and is irreducible and aperiodic (see
\subfig{fig:TASEPApplication}{b}).

\section*{Lifted \TASEP: Bethe ansatz solution}
In order to solve the eigenvalue equation for the transition matrix in
the $N$-particle sector we introduce the following compact notations
\be
\{\boldsymbol{j};a\}=\{j_1,\dots,j_{a-1},\vec{j}_a,j_{a+1}\dots,j_N\}\ .
\ee
Here,  $j_1< \dots < j_N$ are the
positions of the $N$
particles, and $a$ specifies the active particle.
The transition matrix of the lifted \TASEP \eq{equ:LiftedTasepTransition4} then takes the form
\begin{align}
P^\lTASEP(\{\boldsymbol{k};b\}, \{\boldsymbol{j};a\})
&=\alphabar\delta_{a,b+1}\delta_{k_{b}+1,k_{b+1}}\delta_{\boldsymbol{j},\boldsymbol{k}}\nn
&+\alphabar\delta_{b,a}(1-\delta_{k_{b}+1,k_{b+1}})
\delta_{\boldsymbol{j},\boldsymbol{k}+\boldsymbol{e}_b}\nn
&+\alpha\delta_{b,a+1}(1-\delta_{k_{b}+1,k_{b+1}})
\delta_{\boldsymbol{j},\boldsymbol{k}+\boldsymbol{e}_{b}}\nn
&+\alpha\delta_{b,a}\delta_{k_{b}+1,k_{b+1}}
\delta_{\boldsymbol{k},\boldsymbol{j}}\ .
\end{align}
Here we have defined $\boldsymbol{e}_a=(0,\dots,0,1,0,\dots)$, where
the non-zero entry is at position $a$ and
$\delta_{\boldsymbol{j},\boldsymbol{k}}=\prod_{n=1}^N\delta_{j_n,k_n}$.
The left eigenvalue equation for $P^\lTASEP$ then takes the form
\begin{align}
&\EV\psi_a(\boldsymbol{j})=
\bar{\alpha} \psi_{a-1}(\boldsymbol{j})\ \delta_{j_{a},j_{a-1}+1}\nn
&+\bar{\alpha}\psi_a(\dots,j_a-1,\dots)\
(1-\delta_{j_a,j_{a-1}+1})\nn
&+\alpha
\psi_{a+1}(\dots,j_{a+1}-1,\dots)\ (1-\delta_{j_{a+1},j_{a}+1})\nn
&+\alpha \psi_{a}(\boldsymbol{j})\ \delta_{j_{a+1},j_{a}+1}\ .
\label{SG}
\end{align}
As mentioned, when viewed as a non-Hermitian quantum spin chain
\cite{alcaraz1994reaction}, this corresponds to a three-state model
with infinite-range interactions (see Methods). We solve the
eigenvalue equation \fr{SG} by applying a nested Bethe ansatz
\be
\psi_a(\boldsymbol{j})=\sum_{\Perm\in S_N}A_a(\Perm)\prod_{j=1}^N
(z_j)^{j_{\Perm_j}}\ ,
\label{ansatz}
\ee
where $S_N$ is the set of  all permutations of $(1,\dots,N)$ and
$\SET{z_1 \TO z_N}$ are complex spectral parameters that specify a given
solution. We note in passing that we have not attempted to use
the Bethe ansatz to determine the right eigenvectors of the transition
matrix, which in general differ from the left eigenvectors but have
the same eigenvalues. Substituting the Ansatz of \eq{ansatz} into
\eq{SG}
for the probabilities  and assuming that $|j_n-j_{n-1}|>1$
$\forall\ n$, we obtain
\be
A_a(\Perm)\left[\EV-\frac{\alphabar}{z_{\Perm^{-1}_{a}}}\right]=\frac{\alpha}{z_{\Perm^{-1}_
{ a +1}}}A_{a+1}(\Perm)\ .
\ee
These relations express all $A_a(\Perm)$ in terms of only
$A_{1}(\Perm)$
\be
A_a(\Perm)=\left[\prod_{b=1}^{a-1}
\frac{z_{\Perm^{-1}_{b+1}}}{\alpha}\big(\EV-\frac{\alphabar}{z_{\Perm^{-1}_b}}\big)
\right]A_1(\Perm)\ .
\label{equ:interior}
\ee
Additional relations are obtained by considering ``sector boundaries'' \\
(i) \underline{$j_{a-1}+1<j_a=j_{a+1}-1$}.
Here, the eigenvalue equation
\be
\EV\psi_a(\boldsymbol{j})=\alpha\psi_a(\boldsymbol{j})+\alphabar\psi_a(\dots,j_a
-1,\dots)\ ,
\ee
implies the following relations between amplitudes
\begin{align}
\frac{A_a(C_a(\Perm))}{A_a(\Perm)}&
=\frac{z_{\Perm^{-1}_{a+1}}}{z_{\Perm^{-1}_{a}}}T(z_{\Perm^{-1}_{a+1}},z_{\Perm^
{-1}_{a}})\ ,\nn
T(z_1,z_2)&=-\frac{E-\alpha-\alphabar/z_2}{E-\alpha-\alphabar/z_1
} \ ,
\label{equ:RelationAmplitudes}
\end{align}
where $C_a(\Perm)=\big(\dots \Perm^{-1}_{a+1},\Perm^{-1}_a,\dots\big)^{-1}=(a,a+1)\Perm$
is the permutation obtained from $\Perm$ by swapping $a$ with $a+1$.
These can be rewritten as conditions on $A_1(\Perm)$ using \eq{equ:interior}:
\be
\frac{A_1(C_a(\Perm))}{A_1(\Perm)}=T(z_{\Perm^{-1}_{a+1}},z_{\Perm^{-1}_{a}})\ \begin{cases}
1 & \text{ if }a>1\ ,\nn
\frac{z_{\Perm^{-1}_{a+1}}}{z_{\Perm^{-1}_{a}}}& \text{ if } a=1.
\end{cases}
\label{equ:perms}
\ee
(ii) \underline{$j_{a-1}+1=j_a<j_{a+1}-1$}.
Here, the eigenvalue equation
\be
\EV\psi_a(\boldsymbol{j})=
\alphabar \psi_{a-1}(\boldsymbol{j})+\alpha\psi_{a+1}(\dots,j_{a+1}-1,\dots)
\ee
implies the following relations between amplitudes
\be
\frac{A_a(C_{a-1}(\Perm))}{A_a(\Perm)}=-
\frac{\EV-\alpha-\alphabar/z_{\Perm^{-1}_{a-1}}}{\EV-\alpha- \alphabar/z_
{\Perm^{-1}_{a}}}
\frac{\EV-\alphabar/z_{\Perm^{-1}_{a}}}{\EV-\alphabar/z_{\Perm^{-1}_{a-1
} } }\ .
\ee
These can be rewritten as conditions on $A_1(\Perm)$ using \eq{equ:interior}
\be
\frac{A_1(C_{a-1}(\Perm))}{A_1(\Perm)}=T(z_{\Perm^{-1}_{a}},z_{\Perm^{-1}_{a-1}})\ ,\ a\geq 2.
\label{equ:perms2}
\ee
As an arbitary
permutation can be generated from the transpositions $(a,a+1)$ the
conditions of \eqtwo{equ:perms}{equ:perms2} allow us to relate all amplitudes
$A_1(\Perm)$ to $A_1(1,\dots,N)$. The various relations are equivalent to
the following: let $\Perm$ and $\Perm'$ be two permutations related by a single transposition
\be
\Perm'=\Pi_{t_1,t_2}\Perm\ ,
\ee
where $\Pi$ exchanges the entries $\Perm_{t_1}$ and $\Perm_{t_2}$ and we choose
conventions such that we always have $\Perm_{t_1}<\Perm_{t_2}$. Then we have 
\begin{align}
\frac{A_{1}(\Perm)}{A_{1}(\Perm')}=
T(z_{{t_1}},z_{{t_2}})
\times\begin{cases}
\frac{z_{t_1}}{z_{t_2}} & \text{ if } 1\in\{\Perm_{t_1},\Perm_{t_2})\\
1 & \text{ else.} 
\end{cases}
\label{equ:Connection2}
\end{align}
The first case can be viewed as a scattering process between the
active particle and an ordinary particle, while the latter
corresponds to a scattering process involving two ordinary particles. 

All conditions arising from other sector boundaries
(\eg\ three neighboring particles) reduce to
\eqtwo{equ:interior}{equ:Connection2}.

\subsection*{Periodic boundary conditions}
The periodic boundary conditions impose that
\begin{align}
\psi_N(j_1,\dots,j_{N-1},L+1)&=\psi_1(1,j_1,\dots,j_{N-1})\ ,\nn
\psi_1(0,j_2,\dots,j_{N})&=\psi_N(j_2,\dots,j_{N},L).
\end{align}
They give rise to equations that determine the spectral parameters
$z_j$
and the eigenvalue $\EV$
\begin{align}
&z_a^{L-1}=\frac{\alphabar/z_a -\EV}{\alpha}\prod_{b=1}^NT(z_a,
z_b ) \ ,\quad a=1,\dots,N\ ,\nn
  &\prod_{j=1}^N\left(\EV-\alphabar/ z_j \right)
  =\alpha^N\prod_{k=1}^N\frac{1}{z_k}\ . 
\label{equ:BAE}
\end{align}
Explicit diagonalization of $P ^\lTASEP$  for $N\leq 7$ and $L=2N$ confirms the
validity of these equations to machine precision.

\section*{Spectral properties of the lifted \TASEP from the Bethe ansatz}
The spectrum of the logarithm of the transition matrix $\ln(P^\lTASEP)$ can be
determined for large system sizes by analyzing the Bethe equations
\fr{equ:BAE}. Of particular interest are the steady state $\pi$ and the
\quot{excitation gap}, by which we denote the real part of the eigenvalue of
$\ln(P^\lTASEP)$ closest to zero (\cf\ Methods,
\eqtwo{equ:TVDCorrelation1}{equ:TVDCorrelation2}).

For simplicity, we again focus on the case $L=2N$, so that
$\alphacrit=1/2$. It is convenient to
reparametrize the Bethe
equations using
\begin{align}
\beta&=\frac{\alphabar}{\EV-\alpha}\ ,\quad u_a=\frac{2z_a}{\beta}-1\ ,\
\delta=\frac{\beta- \alphabar/ \alpha }{\beta+
\alphabar / \alpha}\ ,\nn
\mu&=\left(\frac{2}{\beta}\right)^L\frac{1-\alpha(1-\beta)}{2\alpha}
\prod_{b=1}^N\frac{u_b-1}{u_b+1}\ .
\end{align}
This maps \eq{equ:BAE} onto
\begin{align}
\big(1-u_a^2\big)^{\frac{L}{2}}&=-\mu(u_a+\delta)\ ,\quad
a=1,\dots,\frac{L}{2}\ ,\nn
\Big(\frac{2\alpha}{1-\alpha(1-\beta)}\Big)^{\frac{L}{2}}&
=\prod_{b=1}^{\frac{L}{2}}(u_b+\delta)\ .
\label{equ:BAEu}
\end{align}
These equations are similar to the ones of the \TASEP with periodic boundary
conditions in that
the equations for $u_a$ can be written as simple polynomial equations
involving a self-consistently determined constant. However, there are
two key differences:
\begin{enumerate}
\item{} The factor $u_a+\delta$ in the equation for $u_a$ makes the
roots lie on a non-trivial contour whereas, for the \TASEP, they lie on a
circle.
\item{} The eigenvalue $\eig$ enters like a root of a nested Bethe equation
rather than being a simple additive function of the roots.
\end{enumerate}

\subsection*{Steady state}
The steady state $\pi$, that by construction is the equal-probability mixture,
has $\EV=1$, which implies that
\be
\beta=1\ ,\quad \delta=2\alpha-1.
\ee
Our equations for $L=2N$ then become
\begin{align}
  \frac{\left(1-u_a^2\right)^{\frac{L}{2}}}{u_a-1+2\alpha}
  &=-\frac{2^L}{2\alpha}\prod_{b=1}^{\frac{L}{2}}
\frac{u_b-1}{u_b+1}\ ,\quad
a=1,\dots,\frac{L}{2}\ ,\nn
\left(2\alpha\right)^{\frac{L}{2}}&=\prod_{b=1}^{\frac{L}{2}}(u_b-1+2\alpha)\ .
\end{align}
Clearly $u_a\to 1$ solves these equations and gives back $\pi$.

\subsection*{Excitation gap for $\alpha=\alphacrit$}
\begin{figure}[tbhp]
\centering
\includegraphics[width=0.8\linewidth]{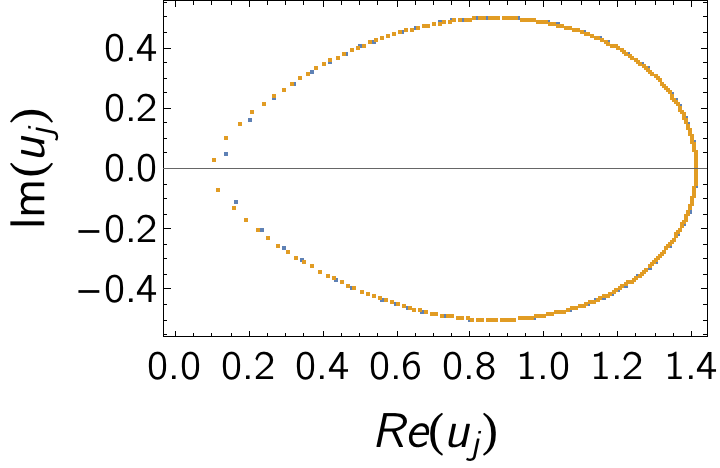}
\caption{Sets of roots $\{u_j|1\leq j\leq L/2\}$ corresponding to the
excited state \fr{ints} for $L=2N$ and $\alpha=\alphacrit = \half$, with
$L=214$ (blue) and
$L=494$ (yellow). For large $L$, the roots approach a non-trivial
contour in the complex plane.}
\label{fig:roots_al05}
\end{figure}

\begin{figure}[ht!]
\includegraphics[width=0.45\textwidth]{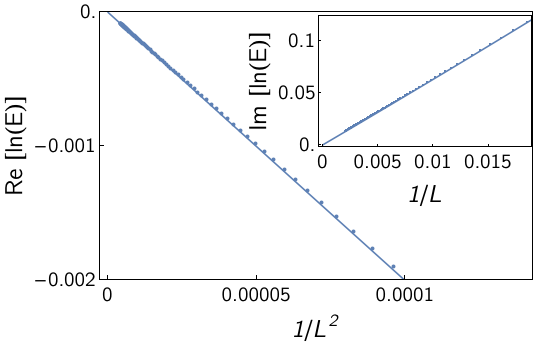}
\caption{Real and imaginary parts of the logarithm of the eigenvalue of
the excited state \fr{ints} as functions of system size for $L=2N$ and
$\alpha=\alphacrit= \half$. The solid lines
are fits described in the main text.}
\label{fig:relogE_al05}
\end{figure}

For simplicity, we fix $L=4m+2$ for integer $m$. We have identified
the structure of solutions of the Bethe equations that give rise to 
low-lying excitations by considering small system sizes $L\leq 14$. We
then follow such solutions for larger values of $L$.
In particular, this leads us to consider the following solution of the
Bethe equations \fr{equ:BAEu}. Out of the $L+2$ roots of
\be
\big(1-u^2\big)^{\frac{L+2}{2}}+\mu_L(u+\delta_L)=0\ ,
\ee
we select a set $\{\bar{u}_1,\dots,\bar{u}_{\frac{L}{2}}\}$ that fulfils
\begin{align}
&\frac{L}{2}\ln\left[\bar{u}_n-1\right]
+\frac{L}{2}\ln\left[-\bar{u}_n-1\right]\nn
&-\ln(-\mu_L)-\ln(\bar{u}_n+\delta_L)=2\pi iI_n\ ,
\end{align}
where 
\be
\{I_n\}=
\{-\frac{L-2}{4},\dots,-1,0,1,\dots\frac{L-2}{4}\} .
\label{ints}
\ee
We find that the root distribution of the corresponding
excited state is of the form shown in \fig{fig:roots_al05}.
The logarithm of the eigenvalue of these states is shown as a function of
system size in \fig{fig:relogE_al05}.
The numerical results are well described by a fit of the form
\be
\ln\big[\EV_1(L)\big]=\frac{a_0}{L}+\frac{a_1}{L^{\frac{3}{2}}}+\frac{a_2}{L^2}
+\frac{a_3}{L^{\frac{5}{2}}}+\frac{a_4}{L^{3}}
+{\cal
O}(L^{-\frac{7}{2}}),
\ee
where $a_0=6.284698i$, $a_1=-0.047459i$, $a_2=-20.0924+6.673286i$,
$a_3=6.04464$ and $a_4=-29.3419$. Interestingly, the real part of
$\ln [E_1(L)]$ appears to scale as $L^{-2}$, while the imaginary part
scales as $L^{-1}$. Some comments are in order. The state \eq{ints} is
the lowest excited state among those we have considered only for
large $L$. This is unusual for Bethe-ansatz solvable models.
Therefore, ${\rm Re}\big(\ln[\EV_1(L)]\big)$ is only an
upper bound for the excitation gap.

\subsection*{Excitation gap for $\alpha\neq \alphacrit$}
\begin{figure}[ht!]
\includegraphics[width=0.4\textwidth]{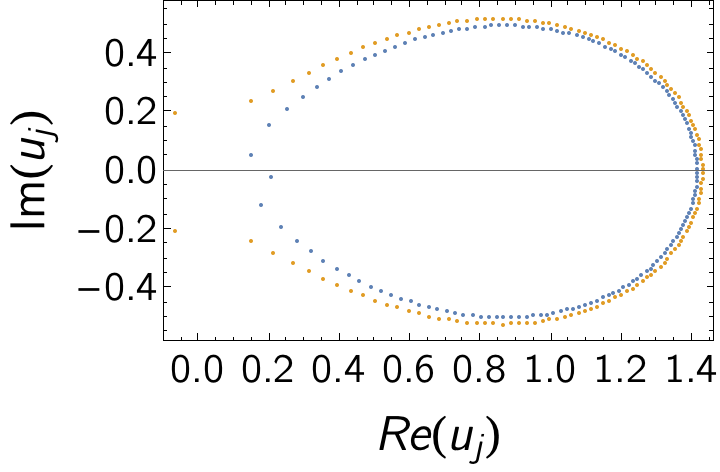}
\caption{Sets of roots $\{u_j|1\leq j\leq L/2\}$ corresponding to the
excited state \fr{ints} for $\alpha=0.9$, $N=L/2$ and $L=272$ (yellow) and
$\alpha=0.4$, $N=L/2$ and $L=290$ (blue). }
\label{fig:roots_al_neq_05}
\end{figure}

\begin{figure}[ht!]
\includegraphics[width=0.45\textwidth]{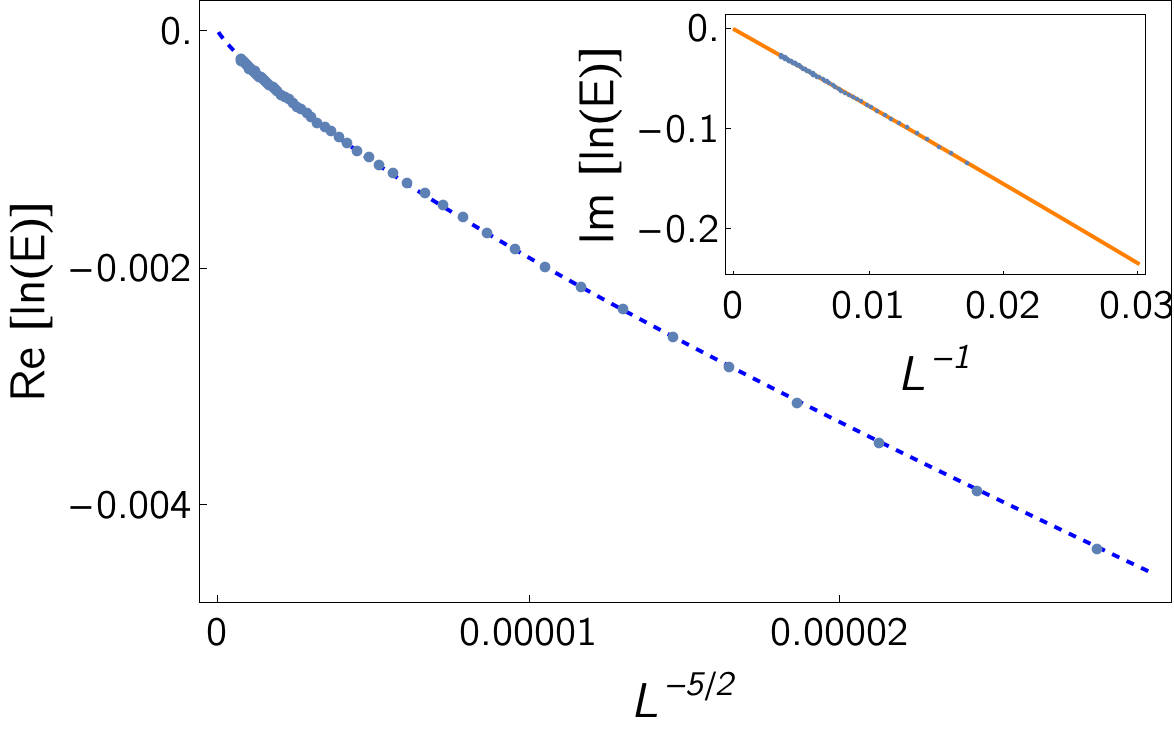}
\caption{Real part of the logarithm of the eigenvalue of
the excited state described in the text for $\alpha=0.4$ as functions
of system size. The solid lines are fits described in the main text.} 
\label{fig:relogE_al04}
\end{figure}

For $L = 2N$,
we have determined the scaling with system size of the eigenvalues of
the solution specified by the same set of integers $\{I_n\}$ in \eq{ints}
for several values of $\alpha\neq 1/2$. This again provides an upper
bound of the excitation gap for large values of $L$. In
\fig{fig:roots_al_neq_05} we show the root distributions for
$\alpha=0.3$, $\alpha=0.4$ and $\alpha=0.9$ respectively.
The logarithm of the eigenvalue of these states is shown as a function of
system size in \fig{fig:relogE_al04}.
Here we fit the excitation gap to the functional form
\begin{align}
\text{Re}\left[\ln\big(\EV_1(L)\big)\right]&=\frac{a_0}{L^\frac{5}{2}}+\frac{a_1}{L^{3}}+\frac{a_2}{L^\frac{7}{2}}+{\cal
O}(L^{4}),\nn
\text{Im}\left[\ln\big(\EV_1(L)\big)\right]&=\frac{b_0}{L}+\frac{b_1}{L^{\frac{3}{2}}}+\frac{b_2}{L^2}+{\cal
O}(L^{\frac{5}{2}}),
\end{align}
where for $\alpha=0.4$, we find
\begin{align}
a_0&=-573.913\ ,\quad
a_1=5651.58\ ,\quad
a_2=-18183.4\ ,\nn
b_0&=-7.54557\ ,\quad
b_1=0.167695\ ,\quad
b_2=-7.61005\ .
\end{align}
For $\alpha=0.9$, the eigenvalue is real for all $N$, with
\begin{align}
a_0&=-19.1347\ ,\quad
a_1=24.5024\ ,\quad
a_2=-1.94545\ .
\end{align}

\section*{Discussion} 

Non-reversible lifted MCMC algorithms such as the lifted \TASEP
are all rooted in reversible Markov chains.
In this paper,  we have taken as our starting point the
one-dimensional lattice reduction of the reversible hard-sphere Metropolis
algorithm, which is nothing but  the celebrated \SSEP, itself a key paradigm for
the stochastic dynamics of interacting particle systems.

The \SSEP has been solved exactly and
its mixing and relaxation times are known to scale differently with system
size~\cite{Lacoin2016detailed,Lacoin_2017_SSEP}. Its $\trel \sim N^3 $
scaling can be viewed as a particle diffusion in
the Edwards--Wilkinson universality class \cite{edwards1982surface}. The
scaling also follows from the proof~\cite{Caputo2010} of the famous Aldous
conjecture which identifies single-particle diffusion with the dynamics
of exclusion
models~\cite{Levin2017}.
We have constructed a second-generation lifting of the \SSEP by inflating
(\quot{lifting}) each
of its configurations $2N$ fold. Because liftings are by design local in space,
it is known~\cite{Chen1999} that the maximal possible speedup (\ie\ the
reduction of mixing and relaxation time scales) scales at best with an exponent
that is half the one of the base model. This corresponds to a
diffusive-to-ballistic speed-up ~\cite{Krauth2021eventchain}.
The best-case inverse-gap
scaling for liftings of the \SSEP is $\sim N^{3/2}$ (up to multiplicative
logarithms). Our MCMC simulations of the lifted \TASEP are indeed
compatible with this optimal scaling behavior for \pullback $\alpha=
\alphacrit$: both the autocorrelation time $\tauauto$ of the structure factor
and the motion of the activity appear to scale a $N^{3/2}$. For $\alpha \neq
\alphacrit$, the same analysis suggests $\tauauto \sim N^{5/2}$.
In contrast, our simulations of the mixing time for a class of compact initial
configurations suggests a scaling of $\tmix \sim N^2$ for $\alpha=\alphacrit$.

As we have shown here,
the lifted \TASEP is exactly solvable by a nested Bethe ansatz in
spite of its non-local structure. The Bethe equations derived in
above yield exact information in particular on the spectrum 
of its transition matrix for large system sizes which
are totally out of reach for other methods. Indeed,
the analysis of these equations for specific families of numerically
exact solutions with $L=2N$ shows that the relaxation time cannot
scale faster than $\sim N^2$ at $\alphacrit$ and $\sim
N^{5/2}$ for $\alpha \neq \alphacrit$. While the latter agrees with our
simulations of $\tauauto$, the former is slower (but still much faster than for
the \SSEP or the \TASEP). A possible explanation of this discrepancy is that
$S(q)$ has for large $N$ negligible overlap with the mode that gives rise to the
$\sim N^2 $ scaling of the inverse gap. We thus once again witness the
exceptional difficulty of estimating the timescales $\tmix$ and
$\trel$ in MCMC, even in the present simplified setting, and the
utility of techniques such as the Bethe ansatz, which
directly access the spectrum of the transition matrix.

\subsection*{Outlook}

Our work raises a number of questions that should be addressed in
future research. First and foremost, a more complete analysis of the spectrum of the
transition matrix in the limit of large system sizes needs to be
carried out, and the scaling behavior of the low-lying
eigenvalues should be derived analytically by recasting the Bethe
equations in terms of as nonlinear integral equations, \cf.
Refs~\cite{Kim1995Bethe,de_Gier2005Bethe,de2006exact,de2008slowest}. Arguably the most important
open question from the point of view of integrability is to determine
the integrable structure of the lifted \TASEP (its R-matrix and L-operator
\cite{korepin1997quantum}). Given the spatially non-local nature of
its transition matrix, it is not obvious how the lifted \TASEP fits
into the existing classification, and it would be interesting to
clarify this. Knowing the integrable structure would be very helpful
in addressing other pressing questions, like whether the partially
asymmetric simple exclusion process has an integrable lifting as well,
and which boundary conditions are compatible with integrability. This
question is natural from the point of view of
Monte-Carlo algorithms and also arises in the context of
boundary-induced phase transitions in the \TASEP. In particular one
could ask whether there is a lifting of the open \TASEP that
reproduces its highly non-trivial steady state \cite{derrida1993exact}
but has a significantly shorter relaxation time. Another point that should
be clarified is whether there is an integrable extension of our model
to several active particles. Finally, we expect the continuum
limit of the lifted \TASEP at vanishing density~\cite{Lei2019} to be
integrable.

Recent decades have witnessed the convergence of rigorous mathematical research
on Markov chains and of the application-driven development of Monte Carlo
algorithms. The present work proceeds further in this direction. The lifted
\TASEP, on the one hand, invites a mathematically rigorous analysis following in
the footsteps of recent work on the \TASEP and the \SSEP. On the other hand,
this new model is precisely the one-dimensional lattice reduction of the
event-chain Monte Carlo algorithm, whose main applications to date have been in
statistical and chemical physics. In this context, the Bethe ansatz has now
entered the game as a  powerful tool, in particular for analyzing complex
spectra. The lifted \TASEP is likely to be the first member of a larger class of
integrable lifted Markov chains, and of non-reversible Monte Carlo algorithms
relevant for applications in the sciences.

\matmethods{
\subsection*{Markov chains} We here collect some basic definitions and
results about Markov chains, lifted Markov chains, and Markov
processes, and discuss the convergence towards the steady state.

\subsubsection*{Basic definition}
We consider finite sample spaces $\Omega$, and Markov chains defined
by (i) a transition matrix $P$,  where $P(x,y)$ gives the conditional probability to
move in one time step to $y$ if at $x$; (ii) an initial distribution
$\pit{t=0}$, generally concentrated on a fixed initial configuration $x_0$. 
The probability distribution at time $t=1,2, \dots$, starting from $x_0$,
is denoted by
$\pit{t} = P^t(x_0,.)$. It satisfies
$\pit{t}_x = \sum_y \pit{t-1}_y P(y,x)$.
An irreducible Markov chain has a unique stationary solution that satisfies
\begin{equation}
\pi_x = \sum_{y \in \Omega} \pi_y P(y,x) \quad \forall x \in \Omega.
\label{equ:GlobalBalance}
\end{equation}
This equation becomes a condition on $\pi$, the \quot{global balance}
condition, in our case where the target distribution is imposed.

The steady state $\pi$ is a row vector, and we refer to it as a left
eigenvector of $P$ with eigenvalue $1$ (that is $\eig \pi = \pi P$, with $\eig
=1$). We call  $\pi$ an equal-probability mixture if all its elements are
identical.
For an irreducible Markov chain that is aperiodic (that is, has no
periodicity in its return to a given configuration $x$), $\pi$ is the limit of
$\pit{t}$ for $t \to \infty$. In the main text, we mention right eigenvectors,
that is, column vectors $\psi^r$ that satisfy $\psi^r = P \psi^r$, and in
particular, the column vector $(1 \TO 1)$, which indicates that the sum of the
possibilities to move away from a configuration $x$ (including back to $x$)
must sum up to one. This means that the matrix is stochastic.
The eigenvalue equation for left eigenvectors reads
\begin{equation}
\eig \psi_x = \sum_{y \in \Omega} \psi_y P(y,x)\quad \forall x \in \Omega\ .
\end{equation}
For an irreducible and aperiodic transition matrix, the eigenvalue $\eig =
1$ is the only one with unit norm.
The detailed-balance condition,
\begin{equation}
\pi_x P(x,y) = \pi_y P(y, x)\quad
\forall x \in \Omega, y \in \Omega\ ,
\label{equ:DetailedBalance}
\end{equation}
implies the global-balance condition, as follows from summing over $y$
and using the conservation of probabilities $\sum_y P(x,y) = 1$.
As discussed in the main text, detailed balance is equivalent to
reversibility, whereas both the non-reversible and reversible Markov
chain satisfy the global-balance condition \eq{equ:GlobalBalance}.

\subsubsection*{Markov chains \emph{vs.} Markov processes}

In the main text, we consider discrete-time Markov chains where at
$t=0,1,2, \dots$,
a single particle is displaced,
modeling  the computer implementation on a serial computer.
The \SSEP and the \TASEP are usually defined as continuous-time
Markov \emph{processes}, where each of the elements of the respective move sets
is
sampled on average once per unit time.
Thus, in the continuous-time \SSEP, on average, $2N$ moves are attempted per
unit time interval
($N$ attempted moves for the continuous-time \TASEP).
The discrete-time and the continuous-time versions are
bijections of each other, with
a rescaling of time scales by a factor of $2N$ for the \SSEP
and a factor of $N$ for the \TASEP,
and with the replacement of equidistant discrete time steps by
Poisson-distributed ones. As an consequence the $\sim N^{5/2}$ relaxation and
mixing-time scales  of the discrete-time \TASEP become $\sim N^{3/2} $
time scales in the continuous-time variant.

\subsubsection*{Lifted Markov chains}

Lifting~\cite{Diaconis2000,Chen1999} connects a Markov chain
with sample space $\Omega$, transition matrix $P$ and steady state
$\pi$ with a lifted Markov chain  $\Pihat$ with $\Omegahat, \Phat, \pihat$
through a mapping $f$ from $\Omegahat $ to $\Omega$. Under this transformation,
$\pi$ must be preserved,
\begin{equation*}
  \pi_v = 
  \sum_{i \in f^{-1}(v)} \pihat_i\ ,
\end{equation*}
where $f^{-1}(v)$ denotes the set of all $i\in\Omegahat$ such that $f(i)=v$.
In addition, the flows must be unchanged
\begin{equation*}
   \underbrace{\pi_v P(v,u)}_{\text{collapsed flow}} = \sum_{i \in f^{-1}(v), j
   \in f^{-1}(u)}
   \overbrace{\pihat_i \Phat(i,j)}^{\text{lifted flow}}\ .
\end{equation*}
In this paper, the lifted sample space $\Omegahat$ is simply the
Cartesian product of the
\quot{collapsed} sample space $\Omega$ with a space $\LCALtilde$ of  lifting
variables, which
correspond to the move set $\LCAL$ or parts of it ($\LCALtilde \subset \LCAL$).
While the lifted sample space is thus enhanced, the set of moves is
smaller, and the lifted transition matrix is sparser. Although the
transition matrix and its lifted version are closely related, the
mixing and relaxation times may be decreased (at most) to their square
roots~\cite{Chen1999,Krauth2021eventchain}. 

Furthermore, in our case all the steady-state probabilities are
constant, which implies that
\begin{equation}
    P(v,u) = \frac{|\Omegahat|}{|\Omega|} \sum_{i \in f^{-1}(v), j
   \in f^{-1}(u)} \Phat(i,j).
\label{equ:SumOverTerms}
\end{equation}
where $|\Omegahat|$ denotes the size of the sample space $\Omegahat$.

\subsubsection*{Convergence}
For an irreducible and aperiodic Markov
chain, $\pit{t}$ converges towards $\pi$ for $t \to \infty$ from any initial
configuration $x_0$. This convergence is
commonly quantified by the worst-case
distance $\dTV(t)$
\begin{equation}
 \dTV(t) = \max_{x_0} || P^t(x_0,.) - \pi || _{\text{TV}},
\end{equation}
where $||\mu  - \nu ||_{\text{TV}}$ denotes the total variation distance (TVD) 
between two distributions $\mu$ and $\nu$: 
\begin{equation}
||\mu  - \nu ||_{\text{TV}} = \frac12 \sum\nolimits_{i \in \Omega} |\mu_i - 
\nu_i|.
\end{equation}
The mixing time $\tmix(\epsilon)$ (which is known for the \SSEP and the \TASEP,
and which we determine using numerical methods for the lifted \TASEP).
It is defined as the earliest time at which $\dTV(t) = \epsilon$, and 
satisfies $\dTV[\ell \tmix(\epsilon)] \le (2 
\epsilon) ^\ell$ for $\ell = 1,2, \dots$. This yields a bound for all times
larger than the mixing time:
\begin{equation}
\dTV(t) \le \expb{- t / \tmix} \quad \text{(for $t > \tmix $)},
\label{equ:TVDMixing}
\end{equation}
where we define $\tmix= \tmix\big(1/(2 \text{e})\big)$.
The mixing time  $\tmix$ is a non-asymptotic timescale with a
strict inequality for finite times given by \eq{equ:TVDMixing}.
Importantly, the mixing time can be larger than the inverse gap (in the
\SSEP, for example $\tmix \sim N^3 \loga N$, whereas the inverse gap scales as
$\trel \sim N^3$. This insight is at the basis of the
modern theory of Markov chains~\cite{AldousDiaconis1986,Levin2017}.

However, for asymptotically large times the worst-case distance
approaches zero as
\begin{equation}
\lim_{t\to \infty} \dTV(t) ^{1/t} = \lambda^*,
\label{equ:TVDCorrelation1}
\end{equation}
where $\lambda^*$ is the largest absolute value of transition matrix
eigenvalues different from $\lambda_1 = 1$ (see
Ref.~\cite[eq. (12.37)]{Levin2017}. Defining the spectral gap by
$\gamma^* = -\ln(\lambda_*)$ and the relaxation time as $\trel =
1 / \gamma^*$, one then has an asymptotic expression
\begin{equation}
 \dTV \sim \expb{-\gamma^* t } \quad \text{(for $t \to \infty $)}\ .
\label{equ:TVDCorrelation2}
\end{equation}

In the \SSEP, for example, the mixing time scales more slowly than the
relaxation time. This leads to a sudden collapse of $\dTV$ (on a time
scale $ \tmix$), as it switches over from a slow decay on the
mixing-time scale (see \eq{equ:TVDMixing}) to a faster decay on the
relaxation-time scale $t_{\rm rel}=1/\gamma^*$. 

The relaxation time remains of relevance in equilibrium (rather than only for the
asymptotic approach to equilibrium). For reversible Markov chains, one has
\begin{equation}
\Var_\pi P^t g \leq  \Var_\pi g \expb{-\lambda_1 t}.
\end{equation}
where $g$ is an arbitrary function on the sample space.

\subsection*{Example transition matrices}
For concreteness, we provide examples for transition
matrices for the standard and the lifted random walks,
as well as the \SSEP and its liftings, the \TASEP and the lifted \TASEP.

\subsubsection*{Random-walk transition matrices (examples)}
Here, we give the transition matrices of the random walk for $L=4$ sites, in the 
basis:
\begin{align*}
1 \equiv\,  & \scalebox{0.8}{\ONE \ZERO \ZERO \ZERO} \quad
2 \equiv\,  \scalebox{0.8}{\ZERO \ONE \ZERO \ZERO} \quad
3 \equiv\,  \scalebox{0.8}{\ZERO \ZERO \ONE \ZERO} \quad
4 \equiv\,  \scalebox{0.8}{\ZERO \ZERO \ZERO \ONE}\ .
\end{align*}
With hard-wall boundary conditions, the case discussed in the main
text, we have
\begin{equation}
P^{\RW}_{\text{walls}}=
\half
\begin{bmatrix}
   1 &  1  & \oo & \oo \\
   1 &  \oo &   1 & \oo  \\
 \oo & 1  &   \oo & 1 \\
 \oo & \oo & 1  & 1   \\
\end{bmatrix} \ .
\end{equation}
Here and in the following the \quot{$\oo$}s stand for zeros.
This transition matrix is irreducible and aperiodic by virtue of the
presence of diagonal terms. It is doubly stochastic
(each of its rows and of its columns sum to one), so that the \steadyconstant.
As $P^{\RW}$ is symmetric and doubly stochastic the associated process
is reversible. 

\subsubsection*{Lifted-random-walk transition matrix (example)}
In the basis
\begin{equation}
\begin{aligned}
1 \equiv\,  & \scalebox{0.7}{\TWO \ZERO \ZERO \ZERO} \quad \!
3 \equiv\,  \scalebox{0.7}{\ZERO \TWO \ZERO \ZERO} \quad \!
5 \equiv\,  \scalebox{0.7}{\ZERO \ZERO \TWO \ZERO} \quad \!
7 \equiv\,  \scalebox{0.7}{\ZERO \ZERO \ZERO \TWO} \\
2 \equiv\,  & \scalebox{0.7}{\FOUR \ZERO \ZERO \ZERO} \quad \!
4 \equiv\,  \scalebox{0.7}{\ZERO \FOUR \ZERO \ZERO} \quad \!
6 \equiv\,  \scalebox{0.7}{\ZERO \ZERO \FOUR \ZERO} \quad \!
8 \equiv\,  \scalebox{0.7}{\ZERO \ZERO \ZERO \FOUR}\ ,
\end{aligned}
\label{equ:BasisLRW}
\end{equation}
the transition matrix of the lifted random walk with hard-wall boundary 
conditions reads
\begin{equation*}
P^{\text{lRW}}_{\text{walls}}=
\begin{bmatrix}
\oo &\oo& \alphabar&\alpha &  \oo &  \oo &  \oo &  \oo \\
\alphabar   &  \alpha &\oo &\oo&  \oo &  \oo &  \oo &  \oo \\
\oo   &\oo &  \oo &  \oo &  \alphabar &  \alpha &  \oo &  \oo \\
\alpha   &  \alphabar &  \oo &  \oo &\oo &\oo&  \oo &  \oo \\
\oo   &  \oo &\oo&\oo &  \oo &  \oo &  \alphabar &  \alpha \\
\oo   &  \oo &  \alpha &  \alphabar &  \oo &  \oo &\oo & \oo\\
\oo   &  \oo &  \oo &  \oo &\oo&\oo &  \alpha &  \alphabar \\
\oo   &  \oo &  \oo &  \oo &  \alpha &  \alphabar &\oo&\oo
\end{bmatrix}.
\end{equation*}
where $\alphabar = 1 - \alpha$.
It is again doubly stochastic, so that the \steadyconstant. As
required by \eq{equ:SumOverTerms}, its $2
\times 2$ blocks sum up to the corresponding single entries in 
$P^\text{RW}_{\text{walls}}$.

\subsubsection*{\SSEP transition matrices (examples)}
For the \SSEP with two
particles  on four sites, we use the basis
\begin{equation}
\begin{aligned}
1 \equiv\,  & \scalebox{0.8}{\ONE \ONE \ZERO \ZERO} \quad
2 \equiv \scalebox{0.8}{\ONE \ZERO \ONE \ZERO} \quad
3 \equiv \scalebox{0.8}{\ONE \ZERO \ZERO \ONE}  \\
4 \equiv\, & \scalebox{0.8}{\ZERO \ONE \ONE \ZERO} \quad
5 \equiv \scalebox{0.8}{\ZERO \ONE \ZERO \ONE} \quad
6 \equiv \scalebox{0.8}{\ZERO \ZERO \ONE \ONE}\ .
\end{aligned}
\label{equ:BasisSSEP}
\end{equation}
With periodic boundary conditions,
the transition matrix of the \SSEP with periodic boundary conditions is
\begin{equation}
P^{\SSEP}_{\text{pbc}}=
\frac{1}{4}
\begin{bmatrix}
   2 &  1  & \oo & \oo &  1  & \oo\\
   1 & \oo &   1 & 1   & \oo &   1\\
 \oo &   1 &   2 & \oo & 1   & \oo\\
 \oo & 1 & \oo & 2   & 1   & \oo\\
   1 & \oo & 1  & 1  & \oo &  1 \\
 \oo &   1 & \oo & \oo & 1 &  2 \\
\end{bmatrix} 
\label{equ:TransitionSSEPpbc}
\end{equation}
while for hard-wall boundary conditions one has instead
\begin{equation}
P^{\SSEP}_{\text{walls}}=
\frac{1}{4}
\begin{bmatrix}
   3 &  1  & \oo & \oo &  \oo  & \oo\\
   1 & 1 &   1 & 1   & \oo &   \oo\\
 \oo &   1 &   2 & \oo & 1   & \oo\\
 \oo & 1 & \oo & 2   & 1   & \oo\\
   \oo & \oo & 1  & 1  & 1 &  1 \\
 \oo &   \oo & \oo & \oo & 1 &  3 \\
\end{bmatrix} \ .
\label{equ:TransitionSSEPWalls}
\end{equation}
Both these transition matrices are doubly stochastic, so that in both
cases the \steadyconstant. Reversibility again follows from the fact
that the transition matrices are doubly stochastic and symmetric.

\subsubsection*{\TASEP transition matrix (example)}

We first discuss the \quot{bi-directional} \TASEP
for two particles on four sites
in the following basis of
$2\binom{4}{2}=12$ configurations, twice as many as for the \SSEP:
\begin{equation}
\begin{aligned}
1& \equiv  \overrightarrow{\scalebox{0.8}{\ONE \ONE \ZERO \ZERO}} \quad
&
2 \equiv  \overleftarrow{\scalebox{0.8}{\ONE \ONE
\ZERO \ZERO}} \\
3& \equiv \overrightarrow{\scalebox{0.8}{\ONE \ZERO \ONE \ZERO}} \quad
&
4 \equiv \overleftarrow{\scalebox{0.8}{\ONE \ZERO \ONE \ZERO}} \\
\vdots& \quad \quad \vdots & \quad  \vdots \\
11& \equiv \overrightarrow{\scalebox{0.8}{\ZERO \ZERO \ONE \ONE}} 
&
12 \equiv \overleftarrow{\scalebox{0.8}{\ZERO \ZERO \ONE \ONE}}
\end{aligned}
\label{equ:BasisTASEPBidirectional}
\end{equation}
The bi-directional \TASEP
rejects moves  that violate the 
exclusion condition (see first row of \eq{equ:MoveTASEP}), but changes the 
overall direction upon collision with a wall (more precisely,
attempting to move the rightmost particle in configuration \quot{$2$}
leads to \quot{$2$}, while moving the leftmost particle leads to \quot{$1$}).
This gives the following transition matrix
\begin{equation}
P^{\TASEP}_{\text{walls}}=
\frac{1}{2}
\begin{bmatrix}
1  &\oo&1  &\oo&\oo&\oo&\oo&\oo&\oo&\oo&\oo&\oo\\ 
1  &1  &\oo&\oo&\oo&\oo&\oo&\oo&\oo&\oo&\oo&\oo\\ 
\oo&\oo&\oo&\oo&1  &\oo&1  &\oo&\oo&\oo&\oo&\oo\\ 
\oo&1  &1  &\oo&\oo&\oo&\oo&\oo&\oo&\oo&\oo&\oo\\ 
\oo&\oo&\oo&\oo&\oo&1  &\oo&\oo&1  &\oo&\oo&\oo\\ 
\oo&\oo&\oo&1  &1  &\oo&\oo&\oo&\oo&\oo&\oo&\oo\\ 
\oo&\oo&\oo&\oo&\oo&\oo&1  &\oo&1  &\oo&\oo&\oo\\ 
\oo&\oo&\oo&1  &\oo&\oo&\oo&1  &\oo&\oo&\oo&\oo\\ 
\oo&\oo&\oo&\oo&\oo&\oo&\oo&\oo&\oo&1  &1  &\oo\\ 
\oo&\oo&\oo&\oo&\oo&1  &\oo&1  &\oo&\oo&\oo&\oo\\ 
\oo&\oo&\oo&\oo&\oo&\oo&\oo&\oo&\oo&\oo&1  &1  \\ 
\oo&\oo&\oo&\oo&\oo&\oo&\oo&\oo&\oo&1  &\oo&1    
\end{bmatrix}.
\label{ptasep}
\end{equation}
While $P^{\TASEP}_{\text{walls}}$ is doubly stochastic and concomitantly
the \steadyconstant, it is non-symmetric and the \TASEP is thus
clearly non-reversible. The $2\times 2$ blocks in \eq{ptasep}
again sum up to the corresponding
entries of $P^{\SSEP}_{\text{walls}}$ (see
\eq{equ:TransitionSSEPWalls}) and the bi-directional \TASEP is thus a
lifting of the \SSEP. 

The transition matrix of the bi-directional \TASEP with periodic boundary
conditions can be written
as a lifting of the \SSEP,
in the basis of \eq{equ:BasisTASEPBidirectional}, with appropriate
flipping probabilities
between the forward and backward sectors.
The periodic boundary conditions allow one to restrict one's attention
to the forward sector.
For two particles on four sites, in the
\quot{forward} basis equivalent to that of \eq{equ:BasisSSEP},
this yields:
\begin{equation}
P^{\TASEP}_{\text{pbc}}=
\half
\begin{bmatrix}
   1 &  1  & \oo & \oo & \oo & \oo\\
 \oo & \oo &   1 & 1   & \oo & \oo\\
 \oo & \oo &   1 & \oo & 1   & \oo\\
 \oo & \oo & \oo & 1   & 1   & \oo\\
   1 & \oo & \oo & \oo & \oo &  1 \\
 \oo &   1 & \oo & \oo & \oo &  1 \\
\end{bmatrix}\quad  .
\label{equ:TASEPHalving}
\end{equation}
This transition matrix is irreducible and aperiodic because of the
presence of non-zero diagonal elements. It is an incomplete lifting of
the \SSEP, because of the absence of backward moves. Nevertheless, as
a doubly stochastic matrix, its \steadyconstant. 

\subsubsection*{Lifted \TASEP transition matrix (example)}
For simplicity, we consider the transition matrix
of the lifted \TASEP with periodic boundary conditions only in the forward
sector.
This leaves us with  $N \binom{L}{N}$ configurations. For two
particles on four sites we have
\begin{align*}
1  \equiv\,& \scalebox{0.8}{\TWO\ONE\ZERO\ZERO}  \quad
2  \equiv \scalebox{0.8}{\ONE \TWO \ZERO \ZERO}  \quad
3 \equiv \scalebox{0.8}{\TWO \ZERO \ONE \ZERO}  \quad
4 \equiv\scalebox{0.8}{\ONE \ZERO \TWO \ZERO} \\ 
5   \equiv\,& \scalebox{0.8}{\TWO\ZERO\ZERO \ONE} \quad
6 \equiv \scalebox{0.8}{\ONE \ZERO \ZERO \TWO}   \quad
7 \equiv \scalebox{0.8}{\ZERO \TWO \ONE \ZERO} \quad
8  \equiv\scalebox{0.8}{\ZERO \ONE \TWO \ZERO} \\ 
9  \equiv\,& \scalebox{0.8}{\ZERO\TWO \ZERO\ONE} \quad
\!\!\!10 \equiv \scalebox{0.8}{\ZERO \ONE  \ZERO \TWO} \quad
\!\!\!11 \equiv \scalebox{0.8}{\ZERO \ZERO  \TWO \ONE} \quad
\!\!\!12 \equiv\scalebox{0.8}{\ZERO \ZERO \ONE \TWO} \ .
\end{align*}
In this basis, the transition matrix of the lifted \TASEP follows
from \eqtwo{equ:LiftedTasepTransition1}{equ:LiftedTasepTransition2} to be
\begin{equation}
P^\lTASEP\!\!= \!\!
\begin{bmatrix}
\alpha&\alphabar&\oo&\oo&\oo&\oo&\oo&\oo&\oo&\oo&\oo&\oo\\
\oo&\oo&\alpha&\alphabar&\oo&\oo&\oo&\oo&\oo&\oo&\oo&\oo\\
\oo&\oo&\oo&\oo&\oo&\oo&\alphabar&\alpha&\oo&\oo&\oo&\oo\\
\oo&\oo&\oo&\oo&\alpha&\alphabar&\oo&\oo&\oo&\oo&\oo&\oo\\
\oo&\oo&\oo&\oo&\oo&\oo&\oo&\oo&\alphabar&\alpha&\oo&\oo\\
\oo&\oo&\oo&\oo&\alphabar&\alpha&\oo&\oo&\oo&\oo&\oo&\oo\\
\oo&\oo&\oo&\oo&\oo&\oo&\alpha&\alphabar&\oo&\oo&\oo&\oo\\
\oo&\oo&\oo&\oo&\oo&\oo&\oo&\oo&\alpha&\alphabar&\oo&\oo\\
\oo&\oo&\oo&\oo&\oo&\oo&\oo&\oo&\oo&\oo&\alphabar&\alpha\\
\alphabar&\alpha&\oo&\oo&\oo&\oo&\oo&\oo&\oo&\oo&\oo&\oo\\
\oo&\oo&\oo&\oo&\oo&\oo&\oo&\oo&\oo&\oo&\alpha&\alphabar\\
\oo&\oo&\alphabar&\alpha&\oo&\oo&\oo&\oo&\oo&\oo&\oo&\oo\\
\end{bmatrix}\ ,
\label{equ:TasepPbcExample}
\end{equation}
where $\alphabar=\alpha-1$. $P^\lTASEP$ is doubly stochastic (so that its
\steadyconstant), irreducible and is aperiodic (as a result of the
presence of non-zero diagonal elements). The transition matrix $P^\lTASEP$
can be partitioned into $2\times 2$ blocks that sum up to the
analogous elements of $P^{\TASEP}_{\text{pbc}}$, as required for a lifting.

\subsection*{Three-particle Bethe ansatz (example)}

For concreteness,  we present a step-by-step
derivation of the Bethe ansatz equations of the lifted \TASEP 
for $N=3$ particles  on an $L$-site lattice with periodic boundary
conditions. The left eigenvalue equations for the transition matrix, in the case
where all three particles are well separated ($j< k-1$, $k<l-1$), take the form
\begin{align}
\eig\psiTwoOneOne{\jmath}{k}{l}&= \alphabar \psiTwoOneOne{j-1}{k}{l} + 
\alpha\psiOneTwoOne{j}{k-1}{l}\ , \nn
\eig\psiOneTwoOne{\jmath}{k}{l}&= \alphabar \psiOneTwoOne{j}{k-1}{l} + 
\alpha\psiOneOneTwo{j}{k}{l-1}\ , \nn
\eig\psiOneOneTwo{\jmath}{k}{l}&= \alphabar \psiOneOneTwo{j}{k}{l-1} + 
\alpha\psiTwoOneOne{j-1}{k}{l}\ . \label{equ:ThreeSector3}
\end{align}

The Bethe ansatz for $\psi$ then reads
\begin{align}
 \psiTwoOneOne{j}{k}{l} &= \ATwoOneOne z_1^j z_2^k z_3^l + \BTwoOneOne 
z_1^j z_2^l z_1^k \!+\! \dots \! + \! \FTwoOneOne z_1^l z_2^k  z_3^j\ ,
\nn
 \psiOneTwoOne{j}{k}{l} &= \AOneTwoOne z_1^j z_2^k z_3^l + \BOneTwoOne 
z_1^j z_2^l z_1^k \!+\! \dots \! + \! \FOneTwoOne z_1^l z_2^k  z_3^j \ ,
 \nn
 \psiOneOneTwo{j}{k}{l} &= \AOneOneTwo z_1^j z_2^k z_3^l + \BOneOneTwo 
z_1^j z_2^l z_1^k \!+\! \dots \! + \! \FOneOneTwo z_1^l z_2^k  z_3^j \ .
 \label{equ:ThreeBethe3}
\end{align}
The three sectors $ \SET{\vv{\bullet} \! \circ \circ}, \SET{\circ 
\vv{\bullet}  \circ}$ and 
$\SET{\circ \circ \! \vv{\bullet}}$ correspond to the position of the active 
particle, while the six coefficients $A, B \TO F$ correspond to the 
permutations of the 
indices $i, j, k$.
Inserting \eq{equ:ThreeBethe3}  into \eq{equ:ThreeSector3}
and comparing coefficients yields 
\begin{align}
\glb \eig - \frac{\alphabar}{z_1} \grb \ATwoOneOne = \frac{\alpha}{z_2} \AOneTwoOne;  
\glb \eig - \frac{\alphabar}{z_1} \grb \BTwoOneOne = \frac{\alpha}{z_3} \BOneTwoOne; 
\dots \nn
\glb \eig - \frac{\alphabar}{z_2} \grb \AOneTwoOne = \frac{\alpha}{z_3} \AOneOneTwo;  
\glb \eig - \frac{\alphabar}{z_3} \grb \BOneTwoOne = \frac{\alpha}{z_2} \BOneOneTwo; 
\dots  \nn
\glb \eig - \frac{\alphabar}{z_3} \grb \AOneOneTwo = \frac{\alpha}{z_1} \ATwoOneOne; 
\glb \eig - \frac{\alphabar}{z_2} \grb \BOneOneTwo = \frac{\alpha}{z_1} \BTwoOneOne;
\dots
\label{equ:CircularThree}
\end{align}
Multiplying these equations
for each of the $A,B, \TO F$ coefficients yields
\begin{equation}
 \prod_{a=1}^3 \glb \eig - \frac{\alphabar}{z_a} \grb =  \frac{\alpha^3}{z_1 z_2
z_3},
\label{equ:ProductFormula3}
\end{equation}
which is identical for $N=3$ to the second expression in \eq{equ:BAE}.

Next, we now consider the eigenvalue equations in the case when two particles
are on neighboring sites and the third particle  is detached:
\begin{align}
\eig\psiTwoOneOne{k-1}{k}{l}&= \alphabar \psiTwoOneOne{k-2}{k}{l} +
\alpha\psiTwoOneOne{k-1}{k}{l}\ , \nn
\eig\psiOneTwoOne{k-1}{k}{l}&= \alphabar \psiTwoOneOne{k-1}{k}{l} + 
\alpha\psiOneOneTwo{k-1}{k}{l-1}\ , \nn
\eig\psiOneTwoOne{j}{l-1}{l}&= \alphabar \psiOneTwoOne{j}{l-2}{l} + 
\alpha\psiOneTwoOne{j}{l-1}{l}\ , \nn
\eig\psiOneOneTwo{j}{l-1}{l}&= \alphabar \psiOneTwoOne{j}{l-1}{l} + 
\alpha\psiTwoOneOne{j-1}{l-1}{l}. \label{equ:OneTouch4}
\end{align}
Substituting \eq{equ:ThreeBethe3} into
\eq{equ:OneTouch4} and comparing coefficients gives
\begin{align}
\frac{\ATwoOneOne}{\CTwoOneOne}& =S_{12};
\frac{\BTwoOneOne}{\DTwoOneOne}=S_{13}; 
\frac{\ETwoOneOne}{\FTwoOneOne}=S_{23}, \nn
\frac{\AOneTwoOne}{\BOneTwoOne}& =S_{23};
\frac{\COneTwoOne}{\EOneTwoOne}=S_{13};
\frac{\FOneTwoOne}{\DOneTwoOne}=S_{21}\ ,
\end{align}
where
\begin{equation}
S_{kl} =  - \frac{z_k}{z_l} \frac{\eig - \alpha - \alphabar / z_l}{\eig - 
\alpha - \alphabar / z_k}.
\end{equation}
Proceeding analogously with the
second and third lines in \eq{equ:OneTouch4} yields:
\begin{align*}
\eig& \glb\! \frac{\AOneOneTwo}{z_2}\!\! + \!\!\frac{\BOneOneTwo}{z_3}\! \grb
-\alphabar \glb\! \frac{\AOneTwoOne}{z_2} \!\! + \!\!\frac{\BOneTwoOne}{z_1} \!
\grb
-\alpha \glb\! \frac{\ATwoOneOne}{z_1 z_2} \!\! + \!\!
\frac{\BTwoOneOne}{z_3z_1} \grb
=0\ , \nn
\eig& \glb\! \frac{\AOneTwoOne}{z_1}\!\! + \!\!\frac{\COneTwoOne}{z_2}\! \grb
-\alphabar \glb\! \frac{\ATwoOneOne}{z_1} \!\! + \!\!\frac{\CTwoOneOne}{z_2}
\!
\grb
-\alpha \glb\! \frac{\AOneOneTwo}{z_3 z_1} \!\! + \!\!
\frac{\CTwoOneOne}{z_3z_2} \grb
=0\ ,
\end{align*}
and five additional  equations that start with $B \TO F$.
The solution to these conditions is:
\begin{align}
\BTwoOneOne =& \frac{z_2}{z_3} S_{32} \ATwoOneOne    \ ,         \nn
\CTwoOneOne =& S_{21}     \ATwoOneOne      \ ,   \nn
\DTwoOneOne =& \frac{z_2}{z_3} S_{31} S_{32}   \ATwoOneOne   \ ,      \nn
\ETwoOneOne =& \frac{z_1}{z_3} S_{21} S_{31}  \ATwoOneOne    \ ,         \nn
\FTwoOneOne =& \frac{z_1}{z_3} S_{21} S_{31} S_{32} \ATwoOneOne\ ,
\label{equ:AToFThree}
\end{align}
which fixes all amplitudes $B \TO F$ in terms of $A$.

Finally, the sector with three adjacent occupied sites
yields the following eigenvalue equations:
\begin{align}
\eig\psiTwoOneOne{j}{j+1}{j+2}&= \alphabar \psiTwoOneOne{j-1}{j+1}{j+2} +
\alpha\psiTwoOneOne{j}{j+1}{j+2}, \label{equ:TwoTouch1}\\
\eig\psiOneTwoOne{j}{j+1}{j+2}&= \alphabar \psiTwoOneOne{j}{j+1}{j+2} +
\alpha\psiOneTwoOne{j}{j+1}{j+2},\label{equ:TwoTouch2} \\
\eig\psiOneOneTwo{j}{j+1}{j+2}&= \alphabar \psiOneTwoOne{j}{j+1}{j+2} +
\alpha\psiTwoOneOne{j-1}{j+1}{j+2}, \label{equ:TwoTouch3}
\end{align}
Of these equations, only \eq{equ:TwoTouch2} leads to a new relation:
\begin{align*}
\gld [E - \alpha] \AOneTwoOne - \alphabar \ATwoOneOne
\grd \frac{z_3}{z_1} +
\gld [E - \alpha] \BOneTwoOne - \alphabar \BTwoOneOne
\grd \frac{z_2}{z_1} + \nn
\gld [E - \alpha] \COneTwoOne - \alphabar \CTwoOneOne
\grd \frac{z_3}{z_2} +
\gld [E - \alpha] \DOneTwoOne - \alphabar \DTwoOneOne
\grd \frac{z_2}{z_1} + \nn
\gld [E - \alpha] \EOneTwoOne - \alphabar \ETwoOneOne
\grd \frac{z_1}{z_2} +
\gld [E - \alpha] \FOneTwoOne - \alphabar \FTwoOneOne
\grd \frac{z_1}{z_3} =0\ .
\end{align*}
Crucially, this condition is identically satisfied as a consequence of
\eqtwo{equ:CircularThree}{equ:AToFThree}.

The periodic boundary conditions:
\begin{equation}
\psiOneOneTwo{j}{k}{L} = \psiTwoOneOne{0}{j}{k},
\end{equation}
give the following relations for the
spectral parameters $ \SET{z_1,z_2,z_3}$:
\begin{equation}
z_a^{L-1}  = \glb \frac{\eig - \alphabar / z_a}{\alpha}      \grb
\prod_{b \ne a = 1}^{3} T(z_a, z_b),
\quad a =1,2,3\ ,
\end{equation}
where
\begin{equation}
T(z_a, z_b) = - \frac{E - \alpha -\alphabar/z_b}{E - \alpha - \alphabar /  z_a}
\end{equation}
(\cf\ \eq{equ:RelationAmplitudes}).

\subsection*{Representation as a non-Hermitian quantum spin chain}
We may consider the transition matrix to induce a continuous time
evolution, which then can be cast in the form of an imaginary-time
Schr\"odinger equation \cite{alcaraz1994reaction}. To do so we define
three basis states $|a\rangle_j$ with $a=0,1,2$ on each site of the
lattice and a basis of operators acting on them 
\be
E_j^{ab}=|a\rangle_j\ {}_j\langle b|\ ,\quad a,b\in\{0,1,2\}.
\ee
The probability distributions of the stochastic process give rise to
states by the map
\be
|\psi\rangle=\sum_{j_1<j_2<\dots<j_N}\sum_{a=1}^N
\psi_a({\boldsymbol{j}})E_{j_1}^{\sigma_1 0}
 E_{j_2}^{\sigma_2 0}\dots E_{j_N}^{\sigma_N 0}|0\rangle\ .
\ee
The master equation takes the form of an imaginary-time
Schr\"odinger equation
\be
\frac{d}{dt}|P(t)\rangle=H|P(t)\rangle\ ,
\ee
where the \quot{Hamiltonian} is given by
\be
H=\sum_{j=1}^N\left[\alphabar +\alpha
  P_{12}\right]\left[E_j^{02}E_{j+1}^{20}+E_j^{12}E_{j+1}^{21}\right]. 
\ee
Here $P_{12}$ is a non-local operator that permutes the active
particle with the regular particle preceeding it:
\be
P_{12}=\sum_{k<j}E_k^{21}\prod_{\ell=k+1}^{j-1}E_\ell^{00}E_j^{12}
+\sum_{j<k}\prod_{\ell=1}^{j-1}E_\ell^{00}E_j^{12}E_k^{21}\prod_{\ell=k+1}^{L}E_\ell^{00}.
\ee

}
\showmatmethods{} 

\acknow{
This work was supported in part by the EPSRC under grant EP/S020527/1 (FHLE)
and the Alexander-von-Humboldt foundation (WK).
We are grateful to A. C. Maggs, C. Monthus, G. Robichon, and M. Staudacher
for helpful discussions.
}

\showacknow{} 

\bibliography{General,LiftedTASEP}

\begin{thebibliography}{41}
\providecommand{\natexlab}[1]{#1}
\providecommand{\url}[1]{\texttt{#1}}
\expandafter\ifx\csname urlstyle\endcsname\relax
  \providecommand{\doi}[1]{doi: #1}\else
  \providecommand{\doi}{doi: \begingroup \urlstyle{rm}\Url}\fi

\bibitem[{Metropolis} et~al.(1953){Metropolis}, {Rosenbluth}, {Rosenbluth},
  {Teller}, and {Teller}]{Metropolis1953}
N.~{Metropolis}, A.~W. {Rosenbluth}, M.~N. {Rosenbluth}, A.~H. {Teller}, and
  E.~{Teller}.
\newblock {Equation of State Calculations by Fast Computing Machines}.
\newblock \emph{J. Chem. Phys.}, 21:\penalty0 1087--1092, 1953.
\newblock \doi{10.1063/1.1699114}.

\bibitem[{Glauber}(1963)]{Glauber1963}
R.~J. {Glauber}.
\newblock {Time-Dependent Statistics of the Ising Model}.
\newblock \emph{J. Math. Phys.}, 4:\penalty0 294--307, 1963.
\newblock \doi{10.1063/1.1703954}.

\bibitem[Creutz(1980)]{Creutz1980HeatBath}
M.~Creutz.
\newblock {Monte Carlo study of quantized SU(2) gauge theory}.
\newblock \emph{Phys. Rev. D}, 21:\penalty0 2308--2315, 1980.
\newblock \doi{10.1103/PhysRevD.21.2308}.
\newblock URL \url{https://link.aps.org/doi/10.1103/PhysRevD.21.2308}.

\bibitem[Geman and Geman(1984)]{Geman1984}
S.~Geman and D.~Geman.
\newblock {Stochastic Relaxation, Gibbs Distributions, and the Bayesian
  Restoration of Images}.
\newblock \emph{IEEE Trans. Pattern Anal. Mach. Intell.}, {PAMI}-6\penalty0
  (6):\penalty0 721--741, 1984.
\newblock \doi{10.1109/tpami.1984.4767596}.
\newblock URL \url{https://doi.org/10.1109/tpami.1984.4767596}.

\bibitem[Diaconis et~al.(2000)Diaconis, Holmes, and Neal]{Diaconis2000}
P.~Diaconis, S.~Holmes, and R.~M. Neal.
\newblock {Analysis of a nonreversible Markov chain sampler}.
\newblock \emph{Ann. Appl. Probab.}, 10:\penalty0 726--752, 2000.
\newblock \doi{10.1214/aoap/1019487508}.

\bibitem[Chen et~al.(1999)Chen, Lovász, and Pak]{Chen1999}
F.~Chen, L.~Lovász, and I.~Pak.
\newblock {Lifting Markov Chains to Speed up Mixing}.
\newblock \emph{Proceedings of the 17th Annual ACM Symposium on Theory of
  Computing}, page 275, 1999.

\bibitem[Spitzer(1970)]{Spitzer1970}
F.~Spitzer.
\newblock {Interaction of Markov processes}.
\newblock \emph{Adv. Math.}, 5\penalty0 (2):\penalty0 246--290, 1970.
\newblock \doi{10.1016/0001-8708(70)90034-4}.
\newblock URL \url{https://doi.org/10.1016/0001-8708(70)90034-4}.

\bibitem[Lacoin(2016)]{Lacoin2016detailed}
H.~Lacoin.
\newblock {The cutoff profile for the simple exclusion process on the circle}.
\newblock \emph{Ann. Probab.}, 44\penalty0 (5):\penalty0 3399--3430, 2016.
\newblock \doi{10.1214/15-aop1053}.
\newblock URL \url{https://doi.org/10.1214/15-aop1053}.

\bibitem[Lacoin(2017)]{Lacoin_2017_SSEP}
H.~Lacoin.
\newblock {The simple exclusion process on the circle has a diffusive cutoff
  window}.
\newblock \emph{Ann. Inst. H. Poincar{\'{e}} Probab. Statist.}, 53\penalty0
  (3):\penalty0 1402--1437, 2017.
\newblock \doi{10.1214/16-aihp759}.

\bibitem[Kapfer and Krauth(2017)]{KapferKrauth2017}
S.~C. Kapfer and W.~Krauth.
\newblock {Irreversible Local Markov Chains with Rapid Convergence towards
  Equilibrium}.
\newblock \emph{Phys. Rev. Lett.}, 119:\penalty0 240603, 2017.
\newblock \doi{10.1103/PhysRevLett.119.240603}.
\newblock URL \url{https://link.aps.org/doi/10.1103/PhysRevLett.119.240603}.

\bibitem[Lei et~al.(2019)Lei, Krauth, and Maggs]{Lei2019}
Z.~Lei, W.~Krauth, and A.~C. Maggs.
\newblock {Event-chain Monte Carlo with factor fields}.
\newblock \emph{Phys. Rev. E}, 99\penalty0 (4), 2019.
\newblock \doi{10.1103/physreve.99.043301}.
\newblock URL \url{https://doi.org/10.1103/physreve.99.043301}.

\bibitem[Bernard et~al.(2009)Bernard, Krauth, and Wilson]{Bernard2009}
E.~P. Bernard, W.~Krauth, and D.~B. Wilson.
\newblock {Event-chain Monte Carlo algorithms for hard-sphere systems}.
\newblock \emph{Phys. Rev. E}, 80:\penalty0 056704, 2009.
\newblock \doi{10.1103/PhysRevE.80.056704}.
\newblock URL \url{http://link.aps.org/doi/10.1103/PhysRevE.80.056704}.

\bibitem[Maggs and Krauth(2022)]{Maggs2022}
A.~C. Maggs and W.~Krauth.
\newblock {Large-scale dynamics of event-chain Monte Carlo}.
\newblock \emph{Physical Review E}, 105\penalty0 (1), January 2022.
\newblock \doi{10.1103/physreve.105.015309}.
\newblock URL \url{https://doi.org/10.1103/physreve.105.015309}.

\bibitem[Peters and de~With(2012)]{Peters_2012}
E.~A. J.~F. Peters and G.~de~With.
\newblock {Rejection-free Monte Carlo sampling for general potentials}.
\newblock \emph{Phys. Rev. E}, 85:\penalty0 026703, 2012.
\newblock \doi{10.1103/PhysRevE.85.026703}.
\newblock URL \url{http://link.aps.org/doi/10.1103/PhysRevE.85.026703}.

\bibitem[{Michel} et~al.(2014){Michel}, {Kapfer}, and {Krauth}]{Michel2014JCP}
M.~{Michel}, S.~C. {Kapfer}, and W.~{Krauth}.
\newblock {Generalized event-chain Monte Carlo: Constructing rejection-free
  global-balance algorithms from infinitesimal steps}.
\newblock \emph{J. Chem. Phys.}, 140\penalty0 (5):\penalty0 054116, 2014.
\newblock \doi{10.1063/1.4863991}.

\bibitem[Bernard and Krauth(2011)]{Bernard2011}
E.~P. Bernard and W.~Krauth.
\newblock {Two-Step Melting in Two Dimensions: First-Order Liquid-Hexatic
  Transition}.
\newblock \emph{Phys. Rev. Lett.}, 107:\penalty0 155704, 2011.
\newblock \doi{10.1103/PhysRevLett.107.155704}.
\newblock URL \url{http://link.aps.org/doi/10.1103/PhysRevLett.107.155704}.

\bibitem[Kampmann et~al.(2021)Kampmann, M\"{u}ller, Weise, Vorsmann, and
  Kierfeld]{Kampmann2021}
T.~A. Kampmann, D.~M\"{u}ller, L.~P. Weise, C.~F. Vorsmann, and J.~Kierfeld.
\newblock {Event-Chain Monte-Carlo Simulations of Dense Soft Matter Systems}.
\newblock \emph{Front. Phys.}, 9:\penalty0 96, 2021.
\newblock \doi{10.3389/fphy.2021.635886}.
\newblock URL \url{https://doi.org/10.3389/fphy.2021.635886}.

\bibitem[Faulkner et~al.(2018)Faulkner, Qin, Maggs, and Krauth]{Faulkner2018}
M.~F. Faulkner, L.~Qin, A.~C. Maggs, and W.~Krauth.
\newblock {All-atom computations with irreversible Markov chains}.
\newblock \emph{J. Chem. Phys.}, 149\penalty0 (6):\penalty0 064113, 2018.
\newblock \doi{10.1063/1.5036638}.
\newblock URL \url{https://doi.org/10.1063/1.5036638}.

\bibitem[Gwa and Spohn(1992{\natexlab{a}})]{Gwa1992six}
L.-H. Gwa and H.~Spohn.
\newblock {Six-vertex model, roughened surfaces, and an asymmetric spin
  Hamiltonian}.
\newblock \emph{Phys. Rev. Lett.}, 68:\penalty0 725--728, 1992{\natexlab{a}}.
\newblock \doi{10.1103/PhysRevLett.68.725}.
\newblock URL \url{https://link.aps.org/doi/10.1103/PhysRevLett.68.725}.

\bibitem[Gwa and Spohn(1992{\natexlab{b}})]{Gwa1992bethe}
L.-H. Gwa and H.~Spohn.
\newblock {Bethe solution for the dynamical-scaling exponent of the noisy
  Burgers equation}.
\newblock \emph{Phys. Rev. A}, 46:\penalty0 844--854, 1992{\natexlab{b}}.
\newblock \doi{10.1103/PhysRevA.46.844}.
\newblock URL \url{https://link.aps.org/doi/10.1103/PhysRevA.46.844}.

\bibitem[Kim(1995)]{Kim1995Bethe}
D.~Kim.
\newblock {Bethe ansatz solution for crossover scaling functions of the
  asymmetric XXZ chain and the Kardar-Parisi-Zhang-type growth model}.
\newblock \emph{Phys. Rev. E}, 52:\penalty0 3512--3524, 1995.
\newblock \doi{10.1103/PhysRevE.52.3512}.
\newblock URL \url{https://link.aps.org/doi/10.1103/PhysRevE.52.3512}.

\bibitem[Derrida(1998)]{derrida1998exactly}
B.~Derrida.
\newblock {An exactly soluble non-equilibrium system: the asymmetric simple
  exclusion process}.
\newblock \emph{Physics Reports}, 301\penalty0 (1-3):\penalty0 65--83, 1998.

\bibitem[Sch{\"u}tz(2001)]{schutz2001exactly}
G.~M. Sch{\"u}tz.
\newblock {Exactly solvable models for many-body systems far from equilibrium}.
\newblock In \emph{Phase transitions and critical phenomena}, volume~19, pages
  1--251. Elsevier, 2001.

\bibitem[Chou et~al.(2011)Chou, Mallick, and Zia]{ChouTASEP2011}
T.~Chou, K.~Mallick, and R.~K.~P. Zia.
\newblock {Non-equilibrium statistical mechanics: from a paradigmatic model to
  biological transport}.
\newblock \emph{Rep. Prog. Phys.}, 74\penalty0 (11):\penalty0 116601, 2011.
\newblock URL \url{http://stacks.iop.org/0034-4885/74/i=11/a=116601}.

\bibitem[Aldous and Diaconis(1986)]{AldousDiaconis1986}
D.~Aldous and P.~Diaconis.
\newblock {Shuffling Cards and Stopping Times}.
\newblock \emph{Am. Math. Mon.}, 93\penalty0 (5):\penalty0 333--348, 1986.
\newblock ISSN 00029890, 19300972.
\newblock URL \url{http://www.jstor.org/stable/2323590}.

\bibitem[Dhar(1987)]{Dhar1987}
D.~Dhar.
\newblock {An exactly solved model for interfacial growth}.
\newblock \emph{Phase Transitions}, 9\penalty0 (1):\penalty0 51, 1987.
\newblock \doi{10.1080/01411598708241334}.

\bibitem[Baik and Liu(2016)]{BaikLiu2016}
J.~Baik and Z.~Liu.
\newblock {TASEP on a Ring in Sub-relaxation Time Scale}.
\newblock \emph{J. Stat. Phys.}, 165\penalty0 (6):\penalty0 1051--1085, 2016.
\newblock ISSN 1572-9613.
\newblock \doi{10.1007/s10955-016-1665-y}.
\newblock URL \url{http://dx.doi.org/10.1007/s10955-016-1665-y}.

\bibitem[Kardar et~al.(1986)Kardar, Parisi, and Zhang]{Kardar1985dynamic}
M.~Kardar, G.~Parisi, and Y.-C. Zhang.
\newblock {Dynamic Scaling of Growing Interfaces}.
\newblock \emph{Phys. Rev. Lett.}, 56:\penalty0 889--892, 1986.
\newblock \doi{10.1103/PhysRevLett.56.889}.
\newblock URL \url{https://link.aps.org/doi/10.1103/PhysRevLett.56.889}.

\bibitem[Derrida and Evans(1999)]{Derrida_1999}
B.~Derrida and M.~R. Evans.
\newblock {Bethe ansatz solution for a defect particle in the asymmetric
  exclusion process}.
\newblock \emph{J. Phys. A: Math. Gen.}, 32\penalty0 (26):\penalty0 4833--4850,
  1999.
\newblock \doi{10.1088/0305-4470/32/26/303}.
\newblock URL \url{https://doi.org/10.1088%2F0305-4470%2F32%2F26%2F303}.

\bibitem[Lei and Krauth(2018)]{Lei2018_TOP}
Z.~Lei and W.~Krauth.
\newblock {Irreversible Markov chains in spin models: Topological excitations}.
\newblock \emph{EPL}, 121:\penalty0 10008, 2018.

\bibitem[Sokal(1997)]{Sokal1997}
A.~Sokal.
\newblock \emph{{Monte Carlo Methods in Statistical Mechanics: Foundations and
  New Algorithms}}, pages 131--192.
\newblock Springer US, 1997.
\newblock ISBN 978-1-4899-0319-8.
\newblock \doi{10.1007/978-1-4899-0319-8_6}.
\newblock URL \url{https://doi.org/10.1007/978-1-4899-0319-8_6}.

\bibitem[Alcaraz et~al.(1994)Alcaraz, Droz, Henkel, and
  Rittenberg]{alcaraz1994reaction}
F.~C. Alcaraz, M.~Droz, M.~Henkel, and V.~Rittenberg.
\newblock {Reaction-diffusion processes, critical dynamics, and quantum
  chains}.
\newblock \emph{Ann. Phys. (N. Y.)}, 230\penalty0 (2):\penalty0 250--302, 1994.

\bibitem[Edwards and Wilkinson(1982)]{edwards1982surface}
S.~F. Edwards and D.~R. Wilkinson.
\newblock {The surface statistics of a granular aggregate}.
\newblock \emph{Proc. R. Soc. A}, 381\penalty0 (1780):\penalty0 17--31, 1982.

\bibitem[Caputo et~al.(2010)Caputo, Liggett, and Richthammer]{Caputo2010}
Pietro Caputo, Thomas Liggett, and Thomas Richthammer.
\newblock {Proof of Aldous' spectral gap conjecture}.
\newblock \emph{Journal of the American Mathematical Society}, 23\penalty0
  (3):\penalty0 831--851, 2010.
\newblock \doi{10.1090/s0894-0347-10-00659-4}.
\newblock URL \url{https://doi.org/10.1090/s0894-0347-10-00659-4}.

\bibitem[Levin and Peres(2017)]{Levin2017}
D.~A. Levin and Y.~Peres.
\newblock \emph{{Markov Chains and Mixing Times, Second Edition}}.
\newblock American Mathematical Society, 2017.

\bibitem[Krauth(2021)]{Krauth2021eventchain}
W.~Krauth.
\newblock {Event-Chain Monte Carlo: Foundations, Applications, and Prospects}.
\newblock \emph{Front. Phys.}, 9:\penalty0 229, 2021.
\newblock ISSN 2296-424X.
\newblock \doi{10.3389/fphy.2021.663457}.
\newblock URL
  \url{https://www.frontiersin.org/article/10.3389/fphy.2021.663457}.

\bibitem[de~Gier and Essler(2005)]{de_Gier2005Bethe}
J.~de~Gier and F.~H.~L. Essler.
\newblock {Bethe Ansatz Solution of the Asymmetric Exclusion Process with Open
  Boundaries}.
\newblock \emph{Phys. Rev. Lett.}, 95:\penalty0 240601, 2005.
\newblock \doi{10.1103/PhysRevLett.95.240601}.
\newblock URL \url{https://link.aps.org/doi/10.1103/PhysRevLett.95.240601}.

\bibitem[de~Gier and Essler(2006)]{de2006exact}
J.~de~Gier and F.~H.~L. Essler.
\newblock {Exact spectral gaps of the asymmetric exclusion process with open
  boundaries}.
\newblock \emph{J. Stat. Mech.}, 2006\penalty0 (12):\penalty0 P12011, 2006.
\newblock \doi{10.1088/1742-5468/2006/12/P12011}.
\newblock URL
  \url{https://iopscience.iop.org/article/10.1088/1742-5468/2006/12/P12011}.

\bibitem[de~Gier and Essler(2008)]{de2008slowest}
J.~de~Gier and F.~H.~L. Essler.
\newblock {Slowest relaxation mode of the partially asymmetric exclusion
  process with open boundaries}.
\newblock \emph{J. Phys. A: Math. Theor.}, 41\penalty0 (48):\penalty0 485002,
  2008.
\newblock \doi{10.1088/1751-8113/41/48/485002}.
\newblock URL \url{https://doi.org/10.1088%2F1751-8113%2F41%2F48%2F485002}.

\bibitem[Korepin et~al.(1997)Korepin, Bogoliubov, and
  Izergin]{korepin1997quantum}
V.~E. Korepin, N.~M. Bogoliubov, and A.~G. Izergin.
\newblock \emph{{Quantum inverse scattering method and correlation functions}},
  volume~3.
\newblock Cambridge University Press, 1997.

\bibitem[Derrida et~al.(1993)Derrida, Evans, Hakim, and
  Pasquier]{derrida1993exact}
B.~Derrida, M.~R. Evans, V.~Hakim, and V.~Pasquier.
\newblock {Exact solution of a 1D asymmetric exclusion model using a matrix
  formulation}.
\newblock \emph{J. Phys. A}, 26\penalty0 (7):\penalty0 1493, 1993.

\end{thebibliography}
\end{document}